\documentclass{pasj00}

\newcommand{\NH}{$N_{\rm H}$}
\newcommand{\Msun}{$M_{\odot}$}
\newcommand{\chisq}{$\chi ^2$/dof}

\begin{document}
\SetRunningHead{H.\ Yamaguchi, K.\ Koyama, and H.\ Uchida}{Fe-Rich Ejecta in RCW~86}
\Received{2011/07/28}
\Accepted{2011/09/05}

\title{Suzaku View of the Supernova Remnant RCW~86: \\
	X-Ray Studies of Newly-Discovered Fe-Rich Ejecta}


\author{%
   Hiroya \textsc{Yamaguchi}\altaffilmark{1,2}
   Katsuji \textsc{Koyama}\altaffilmark{3}
   and
   Hiroyuki \textsc{Uchida}\altaffilmark{3} 
}
 \altaffiltext{1}{Harvard-Smithsonian Center for Astrophysics, 
	60 Garden Street, Cambridge, MA 02138, U.S.A.}
\email{hyamaguchi@head.cfa.harvard.edu}
 \altaffiltext{2}{RIKEN (The Institute of Physical and Chemical Research), 
  2-1 Hirosawa, Wako, Saitama 351-0198, Japan}
 \altaffiltext{3}{Department of Physics, Kyoto University, Kitashirakawa-Oiwake-cho,
	Sakyo, Kyoto 606-8502, Japan}


\maketitle

\begin{abstract}

We report on results of imaging and spectral analysis of the supernova remnant (SNR) 
RCW~86 observed with Suzaku. The SNR is known to exhibit K-shell emission of 
low ionized Fe, possibly originating from supernova ejecta. 
We revealed the global distribution of the Fe-rich plasma in the entire remnant, 
for the first time; the Fe-K emission was clearly detected from the west, north, and south 
regions, in addition to the X-ray brighter shells of southwest and northeast, where the 
presence of the Fe-rich ejecta has already been reported. The spectrum of each region 
is well represented by a three-component model consisting of low- and high-temperature 
thermal plasmas and a non-thermal emission. The lower-temperature component, 
with elemental abundances of near the solar values, likely originates from the forward 
shocked interstellar medium, while the Fe-rich ejecta is described by the hotter plasma. 
From the morphologies of the forward and reverse shocks in the west region, 
the total ejecta mass is estimated to be 1--2\Msun\ for the typical explosion energy 
of $\sim 1 \times 10^{51}$~erg. The integrated flux of the Fe-K emission from the entire 
SNR roughly corresponds to a total Fe mass of about 1\Msun. 
Both of these estimates suggest a Type~Ia supernova origin of this SNR. 
We also find possible evidence of an Fe-rich clump located beyond the forward-shock 
front in the north rim, which is reminiscent of ejecta knots observed in the Tycho and 
Vela SNRs. 

\end{abstract}

\begin{table*}[t]
  \caption{Observation log.}
  \label{tab:log}         
  \begin{center}
    \begin{tabular}{lcccccc}
      \hline
  Region & Obs.\ ID & Obs.\ start date & (RA, Dec)$_{\rm J2000}$ & Exposure & SCI$^{\ast}$ \\
      \hline
  SW & 500004010 & 2006 February 12 & (220.2761, --62.6782) & 101~ks & off \\
  NE & 501037010 & 2006 August 12 & (221.2555, --62.3618) & 60~ks & off \\
  West & 503001010 & 2009 February 2 & (220.2753, --62.4270) & 54~ks & on \\
  North & 503002010 & 2009 January 29 & (220.4956, --62.2074) & 55~ks & on \\
  South & 503003010 & 2009 January 31 & (220.8315, --62.6734) & 55~ks & on \\
  SE/BGD& 503004010 & 2009 February 1 & (221.3859, --62.6710) & 54~ks & on \\
  	\hline
	\multicolumn{6}{l}{$^{\ast}$Spaced-low Charge Injection (Uchiyama et al.\ 2009).}
    \end{tabular}
  \end{center}
\end{table*}

\section{Introduction}
\label{sec:intro}

Making the most of its high spectral sensitivities especially in the hard ($> 5$~keV) 
X-ray band, recent Suzaku observations of extended X-ray sources have continuously 
provided unique results: e.g., the Fe-K line diagnostics of the Galactic center 
plasma (Koyama et al.\ 2007a), the first firm detection of the Fe-K$\alpha$ line from 
SN~1006 (Yamaguchi et al.\ 2008a), and the discoveries of low-abundance elements 
(Cr, Mn, and/or Ni) from the supernova remnant (SNR) Tycho (Tamagawa et al.\ 2009) and 
the Perseus cluster (Tamura et al.\ 2009). The Galactic SNR RCW~86, the remnant of 
SN~185 (e.g., Pisarski et al.\ 1984), is another interesting object where 
the Suzaku performance can be highly utilized, as its spectrum exhibits weak 
K-shell emission from low ionized Fe.

The origin of the Fe-K$\alpha$ emission in RCW~86 had been long-standing issue 
since the discovery. Judging from its center energy of 6.4~keV (consistent with that of 
neutral Fe) together with the presence of hard X-rays, Vink et al.\ (1997) proposed 
that the emission originates from fluorescence caused by supra-thermal electrons. 
However, Bocchino et al.\ (2000), Bamba et al.\ (2000), and Borkowski et al.\ (2001b) 
argued its origin of low ionized ejecta, because the spectrum showed the Fe abundance 
of significantly higher than the solar value. They also revealed that the hard X-ray emission 
does not originate from the supra-thermal bremsstrahlung but non-thermal synchrotron 
emission from relativistic electrons.

The Chandra observation of the southwest (SW) region, the brightest rim of the SNR, 
successfully resolved spatial distribution of the thermal and non-thermal components 
(Rho et al.\ 2002). The non-thermal emission is located at the inward region with respect 
to the optical and soft thermal X-ray filaments of the swept-up interstellar medium (ISM). 
It was suggested, therefore, that the relativistic electrons are accelerated at the reverse 
shock in this region. This is in distinct contrast to typical shell-like SNRs, such as 
SN~1006 and Tycho where non-thermal X-rays are observed close to the blast waves 
(e.g., Koyama et al.\ 1995; Hwang et al.\ 2002). 
In addition, the Fe-K$\alpha$ emission was revealed to be localized at the SNR's interior, 
supporting strongly its origin of the reverse-shocked ejecta. This result was confirmed by 
the Suzaku long-exposure observation (Ueno et al.\ 2007). The authors also found that 
the center energy of the Fe-K$\alpha$ line is slightly higher than that for the neutral Fe, 
suggesting that the emission originates from the high-temperature plasma in 
an extremely low ionization state.

The northeast (NE) region, the second brightest rim, is another site where the detection 
of the Fe-K$\alpha$ emission had been reported (ASCA; Tomida et al.\ 1999). 
However, the following observations by XMM-Newton and 
Chandra failed to detect the emission line (e.g., Vink et al.\ 2006). 
Using Suzaku, Yamaguchi et al.\ (2008b) firmly confirmed the presence of the emission 
and revealed its spatial distribution. It has no spatial correlation with the thermal and 
non-thermal X-ray filaments of the blast wave, but is enhanced at the inner region. 
This is another piece of evidence that the emission originates from the Fe-rich ejecta.

To date, detailed X-ray studies on RCW~86 were mostly limited in the SW and NE 
regions.$\!$\footnote{Spectroscopic results for the north rim were reported by several 
authors (e.g., Bocchino et al.\ 2000; Vink et al.\ 2006), but there is no published report 
regarding the west and south rims.} This is likely because these two rims are particularly 
bright in hard X-rays originating from the synchrotron radiation. 
However, since the preceding studies suggest that the Fe-K$\alpha$ emission is not 
related to the supra- nor non-thermal electrons but is due to the thermal emission from 
the shocked ejecta (e.g., Rho et al.\ 2002; Yamaguchi et al.\ 2008b), 
it should be observed in other X-ray faint regions as well. 
In this paper, we report the first detection of Fe-rich ejecta from the entire remnant, 
including the west, north, and south rims. It should be emphasized that the origin of 
RCW~86 (i.e., Type~Ia or core-collapse supernova) is not revealed yet. 
Since the production of Fe in Type~Ia supernovae is far larger than that of 
core-collapse ones (e.g., Nomoto et al.\ 1984; Iwamoto et al.\ 199), measurement 
of its abundance is one of the essential clues to classify the progenitor type. 
Although RCW~86 is extremely intriguing source for studies of the non-thermal 
phenomena (e.g., Aharonian et al.\ 2009; Helder et al.\ 2009), we here concentrate 
mainly on the thermal emission, and detailed results concerning the non-thermal 
X-rays will be presented in a separate paper (A.\ Bamba et al., in preparation).

We assume the distance to the SNR of 2.8~kpc (e.g., Westerlund 1969; Rosado et al.\ 1996)
throughout the paper. The errors quoted in the text and tables are at the 90\% confidence 
level, and the error bars in the figures are for 1$\sigma$ confidence, unless otherwise stated.

\section{Observation and Data Screening}
\label{sec:obs}

Six pointings were made on RCW~86 with Suzaku in 2006 and 2009 to cover almost 
the entire region of the SNR. Observation logs are found in table~\ref{tab:log}. 
The details of the earlier two observations, SW and NE, are reported by Ueno et al.\ (2007) 
and Yamaguchi et al.\ (2008b), respectively. Here, we present results 
on the other regions obtained with X-ray Imaging Spectrometers (XIS), 
but the SE region was used only as the background data.

The XIS consists of four sensors. Three of them (XIS0, XIS2, and XIS3) are installed with 
front-illuminated (FI) charge-coupled devices (CCDs) whose characteristics are almost 
identical with each other, while the other (XIS1) is installed with a back-illuminated (BI) CCD 
(Koyama et al.\ 2007b). However, the XIS2 was out of operation during the observations 
in 2009 due to damage possibly caused by the impact of a 
micrometeorite.$\!$\footnote{http://heasarc.nasa.gov/docs/suzaku/analysis/abc/node8.html} 
The XISs were operated in normal full-frame clocking mode for all the observations. 
For the data analysis, we used the standard tools of \texttt{HEADAS} version 6.10. 
We reprocessed the public data using \texttt{xispi} software and the latest calibration 
database (CALDB) released on 2011 February 10. The data were reduced in 
accordance with the standard screening 
criteria.$\!$\footnote{http://heasarc.nasa.gov/docs/suzaku/processing/criteria\_xis.html} 
Only grade 0, 2, 3, 4, and 6 events were used in the following analysis.

\section{Results}
\label{sec:res}

\begin{figure}[t]
  \begin{center}
    \FigureFile(65mm,65mm){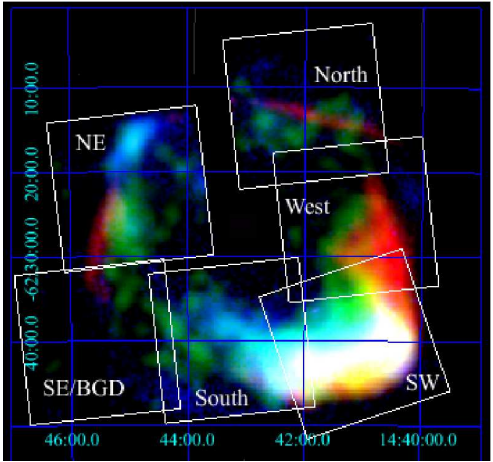}
  \end{center}
  \caption{Three-color XIS image of RCW~86 with a logarithmic scale on the surface 
  	brightness. Red, blue, and green contain emissions from 0.6--2.0~keV (soft), 
	3.0--5.5~keV (hard), and 6.3--6.5~keV (Fe-K), respectively. 
	The coordinates (RA and Dec) refer to epoch J2000.0. 
	The FoVs of the XIS are indicated by the white squares. The data from 
	the three active XIS are combined, but those of the four corners, irradiated by 
	the $^{55}$Fe instrumental calibration sources, are removed from the Fe-K image.
	The exposure and vignetting effect are corrected after the subtraction of the NXB.}
  \label{fig:img_all}
\end{figure}

\begin{figure}[t]
  \begin{center}
    \FigureFile(75mm,75mm){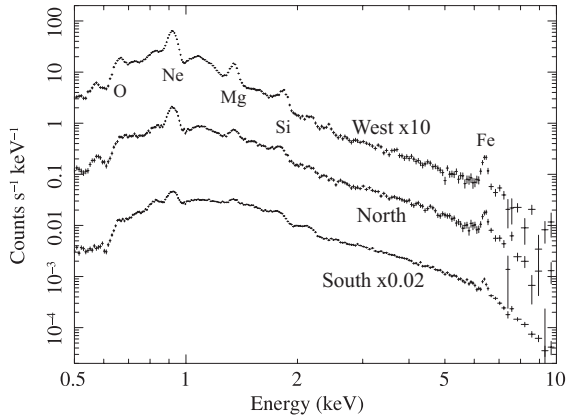}
  \end{center}
  \caption{XIS-FI spectra extracted from each FoV, of which the normalizations are modified.}
  \label{fig:spec_all}
\end{figure}

In figure~\ref{fig:img_all}, we show the three-color XIS image of the SNR, 
where the exposure and vignetting are corrected. Red, blue, and green represent 
0.6--2.0~keV (soft), 3.0--5.5~keV (hard), and 6.3--6.5~keV (Fe-K), respectively. 
As already reported by Ueno et al.\ (2007) and Yamaguchi et al.\ (2008b), the Fe-K 
emission is observed at the SNR's interior of the SW and NE regions. 
In addition, we can see similar Fe-K features in the other field of view (FoV).  
There is no correlation between the spatial distributions of the Fe-K$\alpha$ and hard X-ray 
components, which rejects the possibility of the fluorescence origin for the Fe-K$\alpha$ 
emission as was once suggested by Vink et al.\ (1997). 

We extracted XIS spectra from the entire FoV of each region. 
The result is shown in figure~\ref{fig:spec_all}, where the spectra of 
the two FIs (XIS0 and XIS3) are summed up after the subtraction of non X-ray 
backgrounds (NXB; generated by the \texttt{xisnxbgen} software). 
All the spectra clearly show the Fe-K$\alpha$ lines. 
The large equivalent widths (EW) are observed in the West ($\sim 570$~eV) 
and North ($\sim 480$~eV) regions, while the South region shows smaller 
EW ($\sim 150$~eV) because of considerably large contamination by 
the non-thermal X-rays from the SW rim (e.g., Rho et al.\ 2002). 
We hence present detailed imaging and spectral analysis for the West and 
North regions in the following subsections.

\subsection{West region}
\label{ssec:west}

\begin{figure}[t]
  \begin{center}
    \FigureFile(58mm,58mm){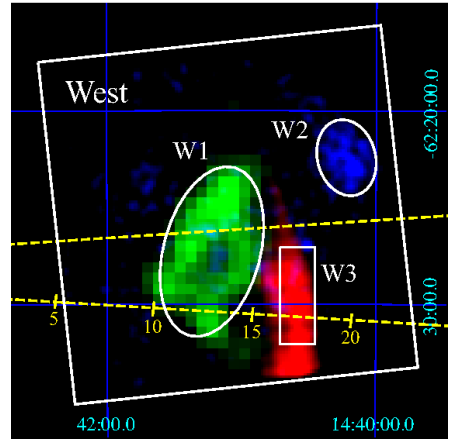}
  \end{center}
  \caption{Three-color image of the West region. The scale is linear. 
  	Red, blue, and green contain emissions from 0.6--2.0~keV, 3.0--5.5~keV, 
	and 6.3--6.5~keV, respectively. 
	The ellipses and rectangle are the regions used in our spectral analysis. 
	The yellow dashed lines with the radius scale (in arcmin) indicate the region 
	where the projection profiles (figure~\ref{fig:proj_w}) are extracted. 
	}
  \label{fig:img_w}
\end{figure}

\begin{figure}[t]
  \begin{center}
    \FigureFile(72mm,72mm){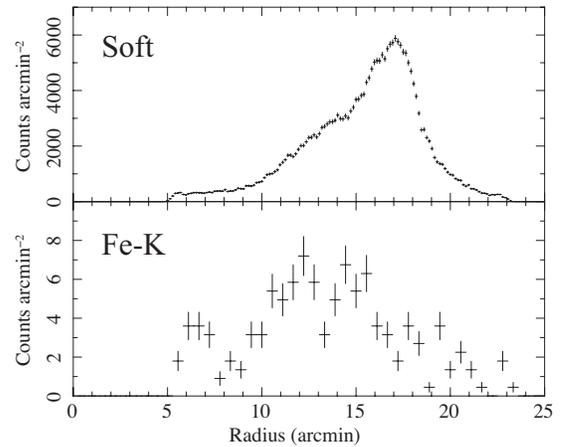}
  \end{center}
  \caption{Projection profiles of the Soft (upper) and Fe-K (lower) band for 
  	the West region. The center of the SNR is assumed to be 
	(RA, Dec)$_{\rm J2000}$ = (220.77, --62.47).}
  \label{fig:proj_w}
\end{figure}

The three-color image of the West region is shown in figure~\ref{fig:img_w}, 
where the corresponding energy ranges are the same as figure~\ref{fig:img_all}. 
The soft and hard band images were binned by each $8 \times 8$~pixels and 
smoothed with a Gaussian kernel of $\sigma$ = 24~pixel ($\simeq 25''$). 
On the other hand, the Fe-K image was binned by each $32 \times 32$~pixels 
(to improve the statistics) with a smoothing Gaussian kernel of $\sigma$ = 96~pixel 
($\simeq 100''$). The soft emission is spatially coincident with the Balmer-dominated 
optical filament (Smith 1997), a signature of a blast wave propagating in 
a tenuous region. Similarly to the SW and NE regions, the Fe-K emission is 
apparently enhanced at the inner region against the soft filament. 
Projection profiles of the soft and Fe-K emissions are given in figure~\ref{fig:proj_w}, 
where (RA, Dec)$_{\rm J2000}$ of (220.77, --62.47) is assumed to be the SNR's 
geometrical center. In addition, we can see the diffuse hard 
emission extending outward to the soft rim. This emission has the radio counterpart 
(Whiteoak \& Green 1996; Dickel et al.\ 2001), and hence likely originates from 
non-thermal X-rays.

\begin{table*}[t]
  \caption{Best-fit parameters for the W1 region.}
  \label{tab:w1} 
  \begin{center}
    \begin{tabular}{llcc}
      \hline
Components & Parameters & Model A & Model B \\
	\hline
Absorption$^{\ast}$ & \NH\ (cm$^{-2}$) & 
	$2.9~(\pm 0.3) \times 10^{21}$ & $2.8~(\pm 0.3) \times 10^{21}$ \\
	
VPSHOCK & $kT_e$ (keV) & 
	$0.46 \pm 0.02$ & $0.46_{-0.02}^{+0.03}$ \\ 
~~(ISM) & $\tau _u$ (cm$^{-3}$~s) & 
	$6.6_{-0.6}^{+0.9} \times 10^{10}$ & $6.8_{-0.6}^{+1.0} \times 10^{10}$ \\
~ & O (solar) & 
	$0.78_{-0.06}^{+0.05}$  & $0.76_{-0.07}^{+0.06}$ \\
~ & Ne (solar) & 
	$1.4 \pm 0.1$ & $1.3 \pm 0.1$ \\
~ & Mg (solar) & 
	$0.71 \pm 0.04$ & $0.72 \pm 0.04$ \\
~ & Fe (solar) & 
	$0.70_{-0.04}^{+0.05}$ & $0.70_{-0.04}^{+0.05}$  \\
~ & $n_e n_{\rm H} dl$ (cm$^{-5}$)$^{\dagger}$ & 
	$6.0_{-0.2}^{+0.1} \times 10^{17}$ & $6.0_{-0.5}^{+0.2} \times 10^{17}$  \\
	
VNEI & $kT_e$ (keV) & 
	\multicolumn{2}{c}{5.0 (fixed)}  \\
~~(Ejecta) & $\tau$ (cm$^{-3}$~s) & 
	\multicolumn{2}{c}{$1.0 \times 10^9$  (fixed)}  \\
~ & $n_e n_{\rm Fe} dl$ (cm$^{-5}$)$^{\dagger}$ & 
	$1.7_{-0.3}^{+0.2}\times 10^{13}$ & $1.7~(\pm 0.3) \times 10^{13}$ \\
	
Non-thermal & $\Gamma$ / $\nu_{\rm rolloff}$ (Hz) & 
	$3.0 \pm 0.1$ & $4.6_{-0.7}^{+0.6} \times 10^{16}$ \\
~ & Norm$^{\ddagger}$ & 
	$5.1_{-0.6}^{+0.4} \times 10^{-5}$ & $4.0~(\pm 0.1) \times 10^{-2}$  \\
  	\hline
\chisq & ~ & 437/280 & 435/280 \\
  	\hline
	\multicolumn{4}{l}{$^{\ast}$\texttt{tbabs} model (Wilms et al.\ 2000).}\\
	\multicolumn{4}{l}{$^{\dagger}$$dl$ is emission depth.}\\
	\multicolumn{4}{l}{$^{\ddagger}$Differential flux (photons~s$^{-1}$~cm$^{-2}$~keV$^{-1}$) 
		at 1~keV in Model~A,}\\
	\multicolumn{4}{l}{~while Flux density in unit of Jy at 1~GHz in Model~B.}\\
    \end{tabular}
  \end{center}
\end{table*}

\begin{figure*}[t]
  \begin{center}
    \FigureFile(76mm,76mm){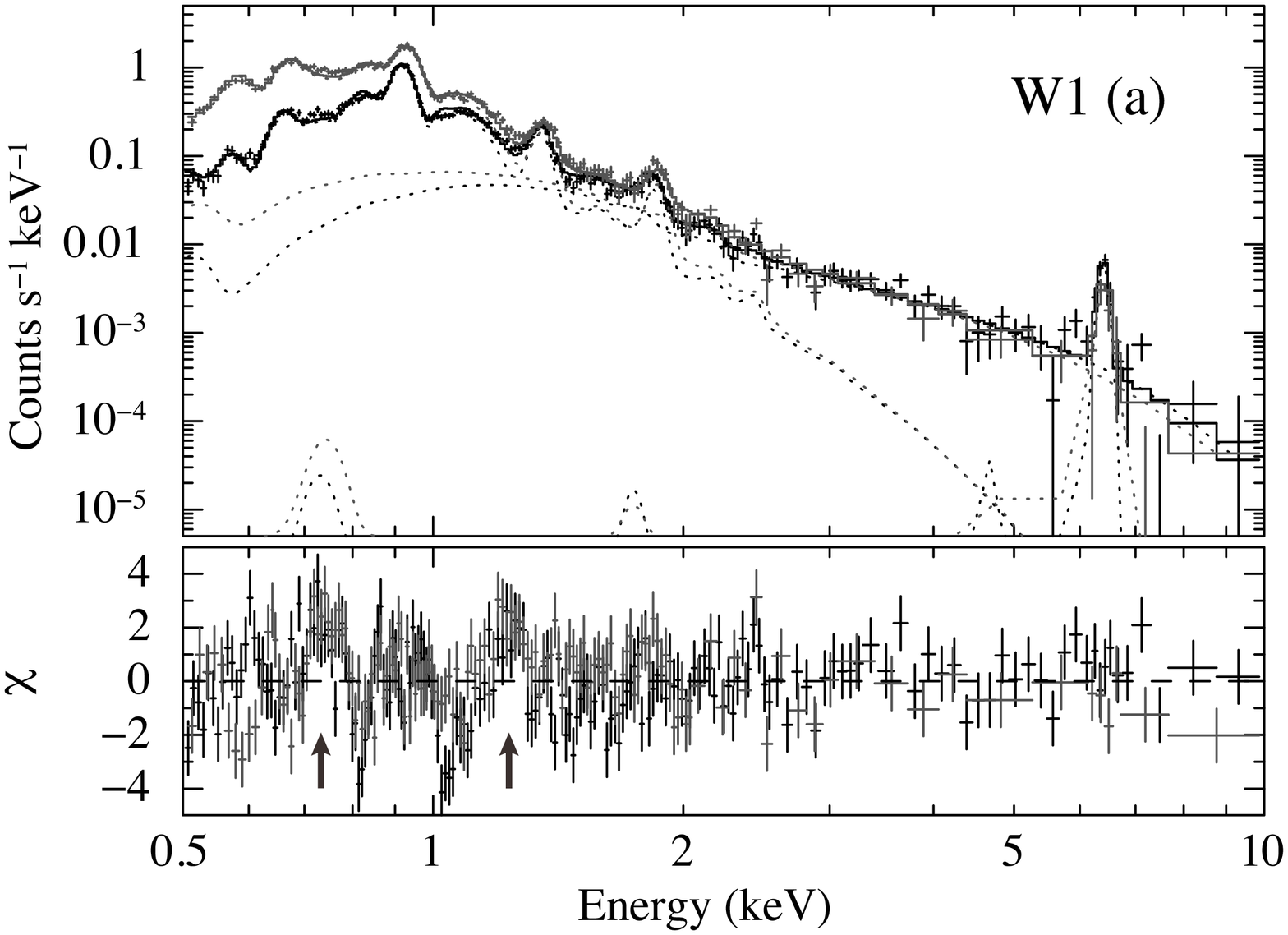}
    \FigureFile(76mm,76mm){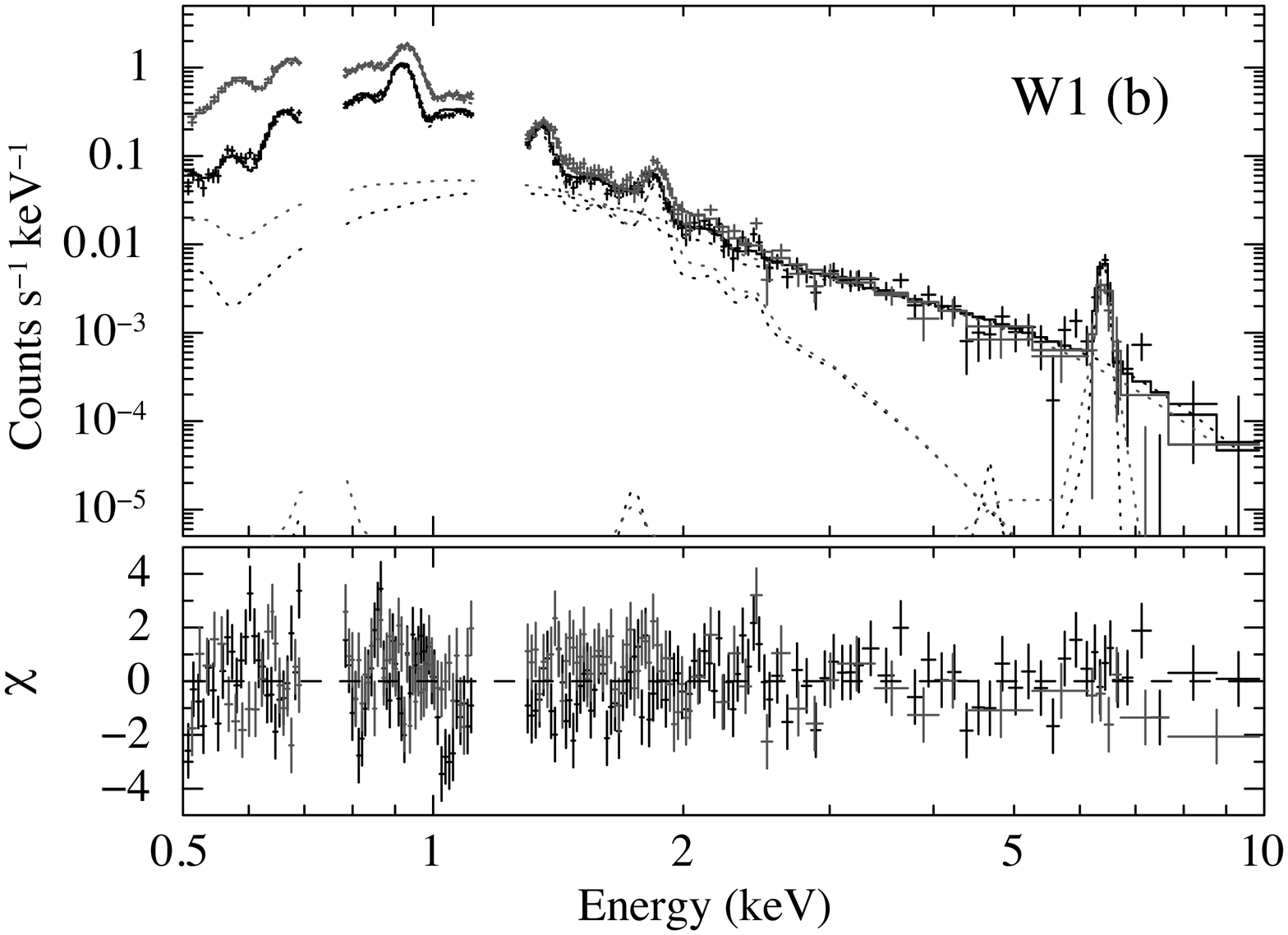}
  \end{center}
  \caption{(a) XIS spectra extracted from the W1 region. The black and gray represent 
  	the FI and BI, respectively. The best-fit model components are given with 
	the dashed lines. The energies where the large residuals are seen 
	(0.73~keV and 1.2~keV) are indicated with black arrows. \ \ 
	(b) Same as (a), but where the problematic energy ranges (0.70--0.78~keV and 
	1.13--1.28) are ignored when fitted (see text for details).}
  \label{fig:spec_w1}
\end{figure*}

We extracted the XIS spectra from three characteristic regions shown in 
figure~\ref{fig:img_w}: W1, W2, and W3, hereafter. 
The background spectra for all the data were taken from the $5' \times 10'$ rectangle 
region in the SE FoV. Although several different background regions were attempted, 
the following results did not essentially change. The background-subtracted spectra 
of the W1 region are shown in figure~\ref{fig:spec_w1}. 
We first analyzed the 4--10~keV band by fitting with a power-law and a Gaussian. 
The center energy of the Fe-K line was determined to be $6403_{-19}^{+21}$~eV, 
suggesting that the ionization state of the Fe is less than +19 (e.g., Makishima 1986; 
Kallman et al.\ 2004).
The obtained EW of  2.7 ($\pm$ 0.5)~keV requires the Fe abundance of at least 
5~solar for a non equilibrium ionization (NEI) plasma with an electron temperature 
of 2.0--10~keV, which is typical range for young SNRs. This is the robust evidence  
that the emission originates from the Fe-rich ejecta.

The result of the imaging analysis (figure~\ref{fig:img_w}) implies that the full-band 
X-ray spectrum of RCW~86 consists of at least three independent components: two 
plasmas responsible to the soft and Fe-K emissions plus a non-thermal component. 
We hence used a plane-parallel shock model (\texttt{vpshock} model; 
Borkowski et al.\ 2001a) to describe the soft thermal component, 
while the Fe-K emission was reproduced by a simpler model of an NEI plasma 
with a single ionization parameter (\texttt{vnei} 
model\footnote{http://space.mit.edu/home/dd/Borkowski/APEC\_nei\_README.txt}).
Free parameters of each component are given in table~\ref{tab:w1}. 
For the soft component, abundances of heavy elements other than O, Ne, Mg, and Fe 
were fixed to the solar values of Anders \& Grevesse (1989). 
For the Fe-K (\texttt{vnei}) component, we made an assumption 
that the plasma consists only of Fe ions and electrons without protons or other heavy 
elements.$\!$\footnote{If we run a fit allowing the Fe abundance to vary freely, 
an extremely large value ($>$100~solar) of the abundance is required. 
This causes difficulty in the determination of both the absolute abundance and 
emission measure, since the bremsstrahlung associated with Fe ions contributes 
significantly to the continuum emission.} 
Since the center energy of the Fe-K line (6403~eV) corresponds to the ionization 
parameter ($\tau$) of $\sim 1 \times 10^9$~cm$^{-3}$~s in the wide temperature 
range of 2--10~keV, we fixed it to this value. 
On the other hand, the electron temperature ($kT_e$) is difficult to be constrained 
because the continuum in the hard X-ray band is likely to be dominated by the 
non-thermal component. Therefore, we assumed $kT_e$ = 5~keV, 
following Rho et al.\ (2002). 
For the non-thermal component, we used a simple power-law (Model~A) or 
an \texttt{srcut} model (Model~B). In the latter case, the energy index in 
the radio band was fixed to $\alpha$ = 0.6 according to Green (2009).
In addition, we allowed for a small offset in the photon energy to the pulse-height 
gain relationship only for the BI data, since the absolute gain of the XIS has 
an uncertainty of a few eV (Koyama et al. 2007b). 
The offset values required for the various regions were about --10~eV, 
within the allowable range of the calibration error.

\begin{figure}[t]
  \begin{center}
    \FigureFile(75mm,75mm){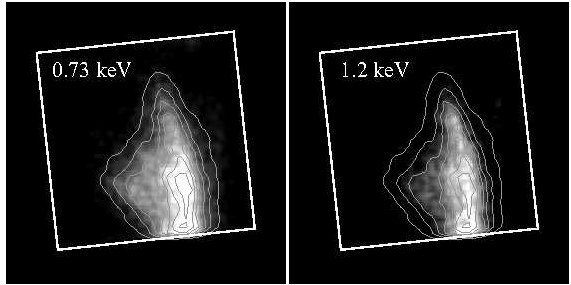}
  \end{center}
  \caption{Gray scale images of the West rim in the energy bands of 0.7--0.78~keV (left) 
  	and 1.13--1.28~keV (right). The intensity contours of the Ne\emissiontype{IX} 
	K$\alpha$ emission are overlaid.}
  \label{fig:img_fe}
\end{figure}

These models and assumptions resulted the best-fit {\tt vpshock} temperature 
and ionization parameter of $kT_e$ = 0.49~($\pm 0.03$)~keV and 
$\tau _u$ = $4.7~(\pm 0.3) \times 10^{10}$~cm$^{-3}$~s, respectively. 
However, the fit was not satisfied with a large 
\chisq\ value of 659/328 (for Model~A) or 673/328 (for Model~B). 
Although we attempted to add another spectral component, 
the fit was not improved significantly with this procedure. 
There are remarkable disagreements between the data and model around 
the energies of $\sim 0.73$~keV and $\sim 1.2$~keV (see figure~\ref{fig:spec_w1}a). 
If we allowed the electron temperature and ionization parameter of the {\tt vnei} 
component to vary freely, the residuals around these energy were reduced. 
This process, however, required a larger $\tau$ value (for the {\tt vnei} component) of 
7.0 (6.8--7.3) $\times 10^9$~cm$^{-3}$~s. This corresponds to the Fe-K center energy of 
$\sim 6.44$~keV, inconsistent with the measured value ($6403_{-19}^{+21}$~eV). 
Therefore, the spectral excesses are unlikely to be associated with the Fe-K emission, 
but seem to originate from the soft thermal (\texttt{vpshock}) component. 
To confirm this, we made the narrow band images in 0.70--0.78~keV and 1.13--1.28~keV 
(figure~\ref{fig:img_fe}) and compared with the intensity 
contours of the Ne\emissiontype{IX} K$\alpha$ line (0.87--0.94~keV), the major component 
of the soft component.
Both the images show apparently the same morphology as the 
Ne\emissiontype{IX} K$\alpha$ emission, supporting that the $\sim 0.73$~keV and 
$\sim 1.2$~keV excesses are associated with the soft emission.

\begin{figure}[t]
  \begin{center}
    \FigureFile(73mm,73mm){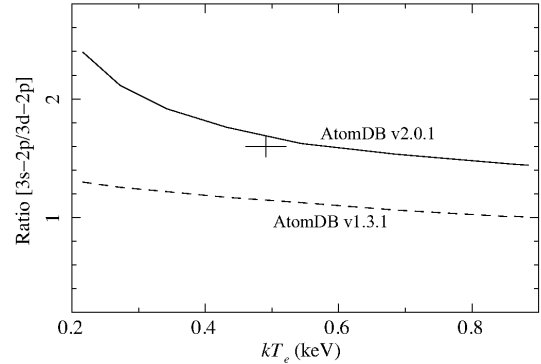}
  \end{center}
  \caption{Emissivity ratio of Fe\emissiontype{XVII} 3s$\to$2p (0.73~keV) to 
  	3d$\to$2p (0.82~keV) lines calculated using the latest AtomDB (version~2.0.0; solid) 
	and previous one (version~1.3.1; dashed), as a function of the electron temperature. 
	The observed ratio from the W1 spectrum is shown as the black cross, 
	where the error bars are for 90\% confidence.}
  \label{fig:ratio}
\end{figure}

\begin{table*}[t]
  \caption{Best-fit parameters for the W2 and W3 regions.}
  \label{tab:w23} 
  \begin{center}
    \begin{tabular}{llcccc}
      \hline
Components & Parameters &  \multicolumn{2}{c}{W2} & \multicolumn{2}{c}{W3} \\
~ & ~ & Model A & Model B & Model A & Model B \\
	\hline
Absorption & \NH\ ($10^{21}$ cm$^{-2}$) & 
	$3.4 \pm 0.3 $ & $3.3_{-0.3}^{+0.2}$ &
	$2.9 \pm 0.2 $ & $2.9_{-0.2}^{+0.3}$ \\
	
VPSHOCK & $kT_e$ (keV) & 
	$0.50_{-0.05}^{+0.08}$ & $0.50_{-0.04}^{+0.05}$ & 
	$0.54_{-0.03}^{+0.02}$ & $0.55 \pm 0.02$ \\ 
~ & $\tau _u$ ($10^{10}$ cm$^{-3}$~s) & 
	$1.5_{-0.7}^{+1.3}$ & $1.7_{-0.3}^{+0.5} $ & 
	$5.4 \pm 0.5$ & $5.4_{-0.4}^{+0.8}$ \\
~ & O (solar) & 
	1$^{\ast}$ & 1$^{\ast}$ & $0.58_{-0.04}^{+0.05}$ & $0.58_{-0.04}^{+0.05}$ \\
~ & Ne (solar) & 
	1$^{\ast}$ & 1$^{\ast}$ & $1.0_{-0.04}^{+0.05}$ & $1.0 \pm 0.1$ \\
~ & Mg (solar) & 
	1$^{\ast}$ & 1$^{\ast}$ & $0.56 \pm 0.03$ & $0.56_{-0.03}^{+0.04}$ \\
~ & Fe (solar) & 
	1$^{\ast}$ & 1$^{\ast}$ & $0.44_{-0.03}^{+0.02}$ & $0.44 \pm 0.03$  \\
~ & $n_e n_{\rm H} dl$ ($10^{17}$ cm$^{-5}$) & 
	$2.0_{-1.0}^{+1.6}$ & $2.0_{-0.3}^{+0.5}$  &
	$15.6_{-0.4}^{+0.2}$ & $16.0_{-0.5}^{+0.6}$ \\
	
VNEI & $n_e n_{\rm Fe} dl$ ($10^{12}$ cm$^{-5}$) & 
	--- & --- &
	$3.7_{-3.5}^{+3.1}$ & $4.0 \pm 3.4$ \\
	
Non-thermal & $\Gamma$ / $\nu_{\rm rolloff}$ (10$^{16}$ Hz) & 
	$3.1 \pm 0.1$ & $3.7_{-0.5}^{+0.7}$ &
	3.0$^{\ast}$ & 4.6$^{\ast}$ \\
~ & Norm$^{\dagger}$ & 
	$1.9 \pm 0.2$ & $0.19_{-0.04}^{+0.05}$ & $0.52_{-0.06}^{+0.05}$ & $0.04 \pm 0.01$ \\
  	\hline
\chisq & ~ & 229/194 & 220/194 & 613/268 & 611/268  \\
  	\hline
	\multicolumn{6}{l}{$^{\ast}$Fixed values.}\\
	\multicolumn{6}{l}{$^{\dagger}$Units are $10^{-4}$~photons~s$^{-1}$~cm$^{-2}$~keV$^{-1}$ 
		(at 1~keV) and Jy (at 1~GHz)}\\
	\multicolumn{6}{l}{~in Model~A and Model~B, respectively}\\
    \end{tabular}
  \end{center}
\end{table*}

\begin{figure*}[t]
  \begin{center}
    \FigureFile(76mm,76mm){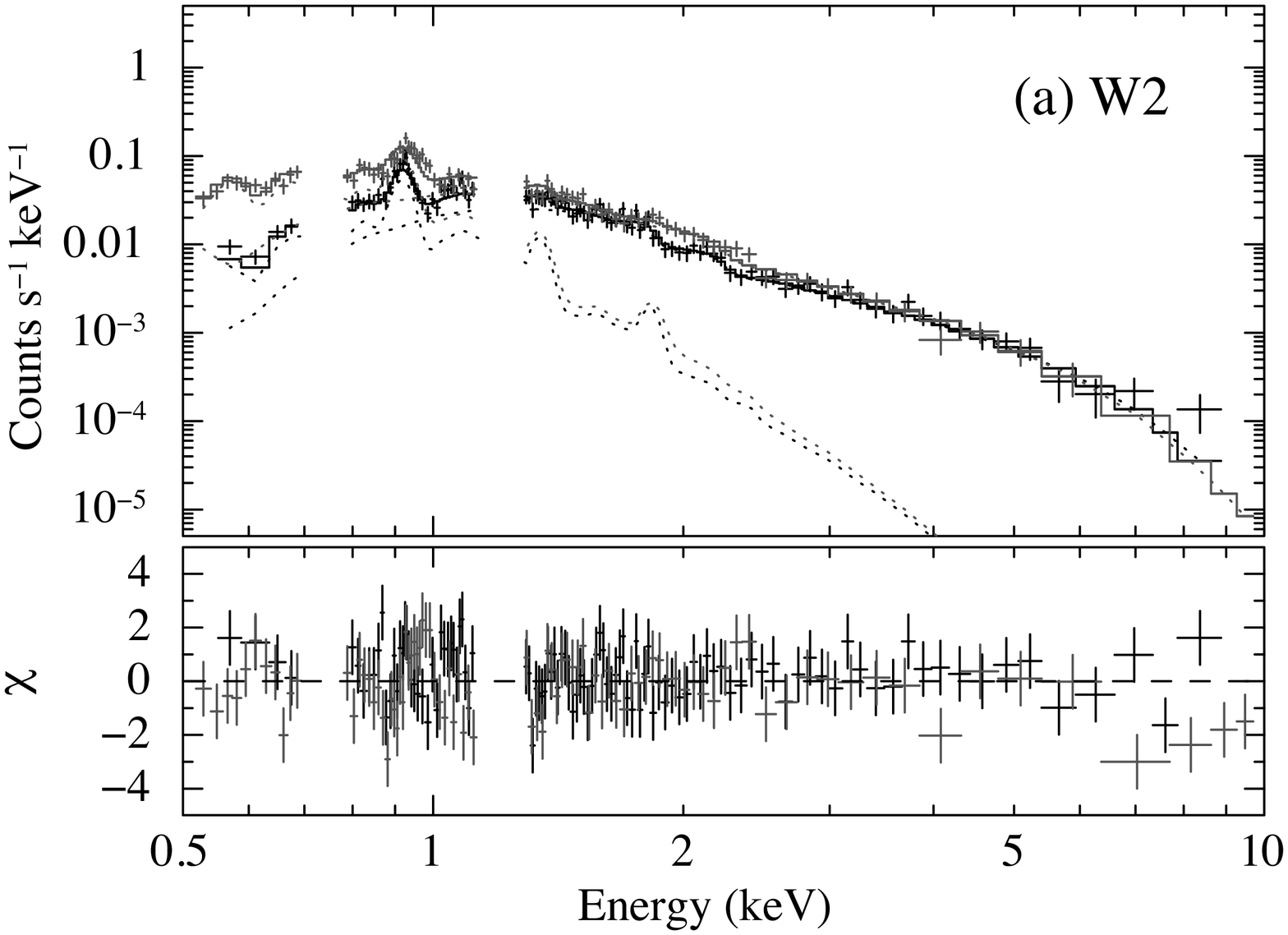}
    \FigureFile(76mm,76mm){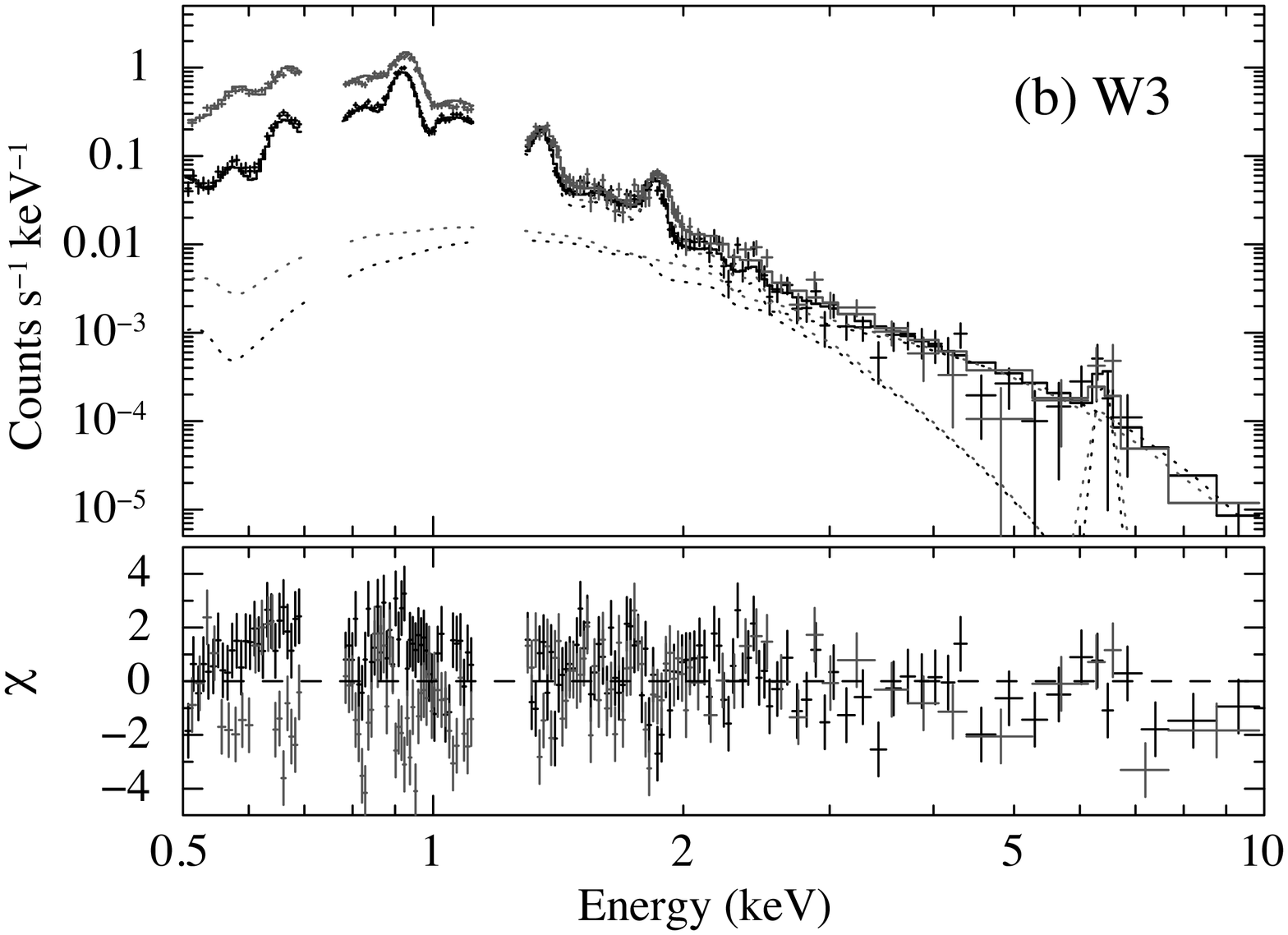}
  \end{center}
  \caption{Same as figure~\ref{fig:spec_w1}, but for the W2 (a) and W3 (b) regions.
  	The energy ranges of 0.70--0.78~keV and 1.13--1.28~keV are ignored.}
  \label{fig:spec_w23}
\end{figure*}

As the origin of the 0.73~keV emission, either O\emissiontype{VII} K-shell lines of 
$n \geq 4$ to 1 or Fe\emissiontype{XVII} L-shell lines of 3s$\to$2p is considered. 
Yamaguchi et al.\ (2008a) found a similar 0.73~keV feature in the spectrum of 
SN~1006, and pointed out that this is likely to be due to the high-level K-shell 
transitions (K$\delta$, K$\epsilon$, K$\zeta$, etc.)~of O\emissiontype{VII} which are 
missing from conventional NEI codes. In SN~1006, an ionization timescale is 
quite low ($\tau \sim 5 \times 10^9$~cm$^{-3}$~s) and an electron temperature is 
moderate ($kT_e \sim 0.5$~keV), so that the contribution of the O\emissiontype{VII} 
K-shell series cannot be ignored. In our case, however, the fluxes of these lines are 
negligibly small compared to those of the O\emissiontype{VIII} emissions because 
of the higher ionization timescale ($\tau_u \sim 5 \times 10^{10}$~cm$^{-3}$~s). 
Therefore, we can eliminate the possibility that the 0.73~keV excess comes from 
O\emissiontype{VII} emissions absent in the plasma code. 
Meanwhile, Katsuda et al.\ (2011) recently suggested that charge-exchange process 
between O\emissiontype{VIII} ions and neutrals can induce the strong emissions of 
O\emissiontype{VII} $n \geq 4$ to 1 transitions. They comprehensively analyzed Suzaku 
and XMM-Newton data of Cygnus Loop and found that the 0.73~keV flux is enhanced 
in several rim regions. However, the charge exchange emissions should be (if any) 
observed only in the outermost rims (Katsuda et al.\ 2011; Lallement 2004), 
and hence unlike in our case.

We thus associate the 0.73~keV feature that we see in the RCW~86 spectrum with 
the Fe\emissiontype{XVII} L-shell emissions. As argued in some previous works 
(e.g., N103B: van der Heyden et al.\ 2002; SNR~0509--67.5: Warren \& Hughes 2004), 
the emissivity ratio between the Fe\emissiontype{XVII} 3s$\to$2p ($\sim 730$~eV) to 
3d$\to$2p ($\sim 820$~eV) transitions is uncertain, which often causes a failure to 
reproduce the data around these energies. 
In figure~\ref{fig:ratio}, we show the 3s--2p/3d--2p emissivity ratios calculated 
using AtomDB\footnote{http://www.atomdb.org/} version~1.3.1 (which is involved 
in the \texttt{vnei} model we used) and recently-updated version~2.0.0. The updated 
database predicts the ratio about twice higher than that of the previous one. 
The primary reason of this change is that the updated one takes into account 
the dielectronic recombination process, in addition to the electron-impact excitation, 
as the source of the L-shell line emissions. 
In the figure, we also plot the 3s--2p/3d--2p flux ratio determined from the W1 spectra. 
This was derived by fitting the 0.5--1.0~keV spectra with an ad hoc model 
consisting of a \texttt{vnei} component with the abundance of Fe fixed to zero and 
three Gaussians at 0.73, 0.82, and 0.90~keV (for the Fe L-shell blends). The obtained 
ratio of $1.62 \pm 0.09$ is almost consistent with the value expected by the updated 
atomic code at $kT_e \sim 0.5$~keV. Therefore, the excess seen around 0.73~keV is 
reasonably explained by the uncertainty of the Fe\emissiontype{XVII} emissivities in 
the currently-distributed NEI model.

Regarding the excess of the data at 1.2~keV, Rho et al.\ (2002) reported 
the presence of a similar spectral feature in the Chandra data of RCW~86 
SW rim. They argued that this was possibly due to the Fe\emissiontype{XVII} blend of 
$n \geq 6$ to 2 that were missing from an NEI code they used. 
Although the atomic database that we used in our analysis includes the electron 
excitations in the Fe\emissiontype{XVII} ions up to the level of $n = 7$, 
the excess at this energy still remains. The origin of this feature is therefore debatable.  
Nevertheless, it is worth noting that the positive correlation between the 1.2~keV flux
and Fe abundance has been confirmed by the spatially-resolved spectral analysis on 
the Cygnus Loop (H.\ Uchida et al., in preparation, where the same plasma code is used). 
We consider, therefore, that the 1.2~keV feature is also related to the Fe\emissiontype{XVII} 
L-shell emissions, or the Ni L-shell blend would be an alternative possibility. 
Ignoring the energy ranges of 0.70--0.78~keV and 1.13--1.28~keV (figure~\ref{fig:spec_w1}b), 
we obtained the best-fit parameters given in table~\ref{tab:w1}. Although the \chisq\ value 
($\sim 1.6$) is still unacceptable from a statistical point of view, we did not apply further 
complicated models.

\begin{table*}[t]
  \caption{Best-fit parameters for the N1 and N2 regions.}
  \label{tab:n12} 
  \begin{center}
    \begin{tabular}{llcccc}
      \hline
Components & Parameters &  \multicolumn{2}{c}{N1} & \multicolumn{2}{c}{N2} \\
~ & ~ & Model A & Model B & Model A & Model B \\
	\hline
Absorption & \NH\ ($10^{21}$ cm$^{-2}$) & 
	$5.3 \pm 0.3$ & $5.0_{-0.6}^{+0.4}$ &
		$5.4_{-0.2}^{+0.6} $ & $5.5_{-0.4}^{+0.2}$ \\
	
VPSHOCK & $kT_e$ (keV) & 
	$0.27_{-0.07}^{+0.02}$ & $0.34_{-0.05}^{+0.08}$ & 
		$0.27_{-0.06}^{+0.02}$ & $0.27 \pm 0.03$ \\ 
~ & $\tau _u$ ($10^{10}$ cm$^{-3}$~s) & 
	$7.9_{-1.3}^{+1.9}$ & $5.0_{-1.8}^{+1.9} $ & 
		$4.5_{-0.5}^{+0.4}$ & $4.6 \pm 0.9$ \\
~ & $n_e n_{\rm H} dl$ ($10^{17}$ cm$^{-5}$) & 
	$8.4_{-2.7}^{+7.2}$ & $4.8_{-2.1}^{+2.6}$  &
		$20.7_{-2.7}^{+19.7}$ & $21.9_{-2.4}^{+9.5}$ \\
	
VNEI  & $n_e n_{\rm Fe} dl$ ($10^{13}$ cm$^{-5}$) & 
	$1.0 \pm 0.3$ & $1.1 \pm 0.3$ &
		$1.0 \pm 0.4$ &  $1.0 \pm 0.5$ \\
			
Non-thermal & $\Gamma$ / $\nu_{\rm rolloff}$ (10$^{16}$ Hz) & 
	$3.2 \pm 0.1$ & $3.4_{-0.5}^{+0.6}$ &
		$3.2 \pm 0.1$ & $3.3_{-0.6}^{+0.8}$ \\
~ & Norm$^{\ast}$ & 
	$1.1 \pm 0.1$ & $0.11_{-0.03}^{+0.04}$ &
		$1.2_{-0.1}^{+0.2}$ & $0.13 \pm 0.04$ \\
  	\hline
\chisq & ~ & 351/213 & 349/213 & 174/166 & 173/166  \\
  	\hline
	\multicolumn{6}{l}{$^{\ast}$Units are $10^{-4}$~photons~s$^{-1}$~cm$^{-2}$~keV$^{-1}$ 
		(at 1~keV) and Jy (at 1~GHz)}\\
	\multicolumn{6}{l}{~in Model~A and Model~B, respectively}\\
    \end{tabular}
  \end{center}
\end{table*}

\begin{figure}[t]
  \begin{center}
    \FigureFile(58mm,58mm){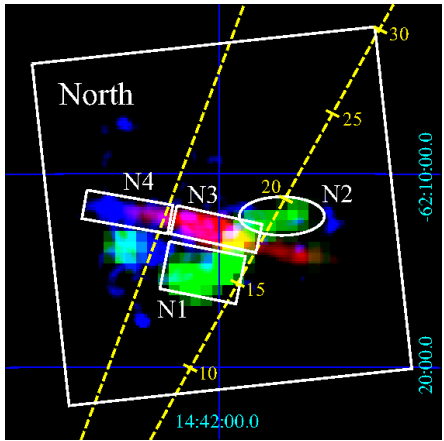}
  \end{center}
  \caption{Same as figure~\ref{fig:img_w}, but for the North rim.
  	The regions used in the spectral analysis are indicated with 
	the ellipse and rectangles. The yellow dashed lines show the 
	region where the projection profiles (figure~\ref{fig:proj_n}) are extracted. }
  \label{fig:img_n}
\end{figure}

We next analyzed the spectra of W2 and W3 regions shown in figure~\ref{fig:spec_w23}. 
The problematic energy ranges of 0.70--0.78~keV and 1.13--1.28~keV were excluded 
from these spectra as well.  
Since the W2 spectrum exhibits no feature of the Fe-K emission, we did not include 
the hotter \texttt{vnei} component. Moreover, the flux of the soft thermal component 
is too weak to obtain a reasonable fit with a lot of free parameters, we fixed all the 
elemental abundances of this component to unity. For the W3 spectrum, 
all the parameters listed in table~\ref{tab:w23} were allowed to vary freely at first. 
The best fit was, however, obtained with an extremely steep slope of the non-thermal 
component, $\Gamma = 4.5$ in Model~A or $\nu_{\rm rolloff} = 6 \times 10^{15}$~Hz 
in Model~B. This is unrealistic because the extrapolated radio flux is required 
to be $\sim 3.0$~Jy at 1~GHz, more than one order of magnitude higher than that of 
the W2 region, resulting considerable disagreement with the radio observation 
(Dickel et al.\ 2001). Therefore, we fixed the slope and roll-off frequency as those 
obtained in the W1 region. The results are given in table~\ref{tab:w23}. 
The relatively large \chisq\ values for the W3 spectrum are mainly caused by 
the inconsistency between the FI and BI data below 1~keV. 
This is probably due to  incomplete calibration information for the contamination 
layer on the optical blocking filter in front of the CCDs (Koyama et al.\ 2007b).

\begin{figure}[t]
  \begin{center}
    \FigureFile(72mm,72mm){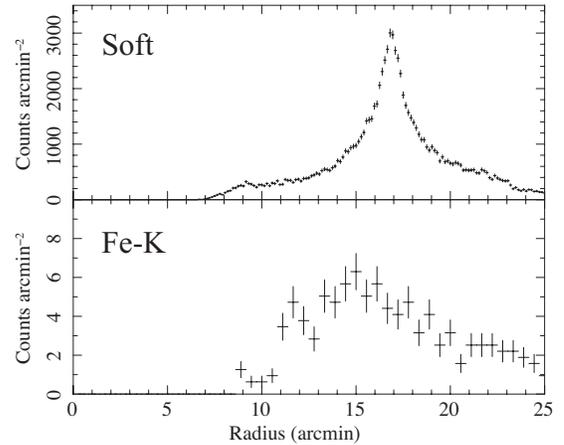}
  \end{center}
  \caption{Same as figure~\ref{fig:proj_w}, but for the North region.}
  \label{fig:proj_n}
\end{figure}

\subsection{North region}
\label{ssec:north}

\begin{table*}[t]
  \caption{Best-fit parameters for the N3 and N4 regions.}
  \label{tab:n34} 
  \begin{center}
    \begin{tabular}{llcccc}
      \hline
Components & Parameters &  \multicolumn{2}{c}{N3} & \multicolumn{2}{c}{N4} \\
~ & ~ & Model A & Model B & Model A & Model B \\
	\hline
Absorption & \NH\ ($10^{21}$ cm$^{-2}$) & 
	$4.6_{-0.4}^{+0.6} $ & $4.5_{-0.4}^{+0.6}$ &
		$4.5_{-0.7}^{+0.6} $ & $4.0_{-0.9}^{+0.6} $ \\
	
VPSHOCK & $kT_e$ (keV) & 
	$0.36_{-0.07}^{+0.05}$ & $0.38 \pm 0.06$ & 
		$0.31_{-0.05}^{+0.11} $ & $0.38_{-0.07}^{+0.14}$ \\ 
~ & $\tau _u$ ($10^{10}$ cm$^{-3}$~s) & 
	$5.0_{-1.1}^{+1.2}$ & $4.6_{-1.0}^{+1.1}$ & 
		$5.4_{-2.5}^{+3.4}$ & $3.7_{-1.2}^{+2.3} $ \\
~ & O (solar) & 
	$1.2 \pm 0.2$ & $1.2_{-0.1}^{+0.2}$ &
		$1.3_{-0.2}^{+0.3}$ & $1.1_{-0.2}^{+0.3}$ \\
~ & Ne (solar) & 
	$1.7 \pm 0.2$ & $1.6 \pm 0.2$ &
		$1.9 \pm 0.5$ & $1.8_{-0.3}^{+0.6}$ \\
~ & $n_e n_{\rm H} dl$ ($10^{17}$ cm$^{-5}$) & 
	$9.6_{-3.3}^{+8.9}$ & $9.1_{-3.2}^{+6.5}$  &
		$7.3_{-4.4}^{+8.0}$ & $4.6_{-2.8}^{+4.9}$ \\
	
VNEI & $n_e n_{\rm Fe} dl$ ($10^{12}$ cm$^{-5}$) & 
	$3.9_{-3.4}^{+3.6}$ & $4.8_{-3.4}^{+3.6}$ &
		$2.6_{-2.6}^{+4.3}$ & $2.6_{-2.6}^{+4.4}$ \\
	
Non-thermal & $\Gamma$ / $\nu_{\rm rolloff}$ (10$^{16}$ Hz) & 
	$3.2 \pm 0.1$ & $3.3_{-0.6}^{+0.8}$ &
		$3.2 \pm 0.1$ & $3.1_{-0.5}^{+0.7}$ \\
~ & Norm$^{\ast}$ & 
	$1.5 \pm 0.1$ & $0.15_{-0.04}^{+0.06}$ &
		$2.2 \pm 0.2$ & $0.23_{-0.01}^{+0.07}$ \\
  	\hline
\chisq & ~ & 297/253 & 299/253 & 227/230 & 225/230  \\
  	\hline
	\multicolumn{6}{l}{$^{\ast}$Units are $10^{-4}$~photons~s$^{-1}$~cm$^{-2}$~keV$^{-1}$ 
		(at 1~keV) and Jy (at 1~GHz)}\\
	\multicolumn{6}{l}{~in Model~A and Model~B, respectively}\\
    \end{tabular}
  \end{center}
\end{table*}

\begin{figure*}[t]
  \begin{center}
    \FigureFile(43mm,43mm){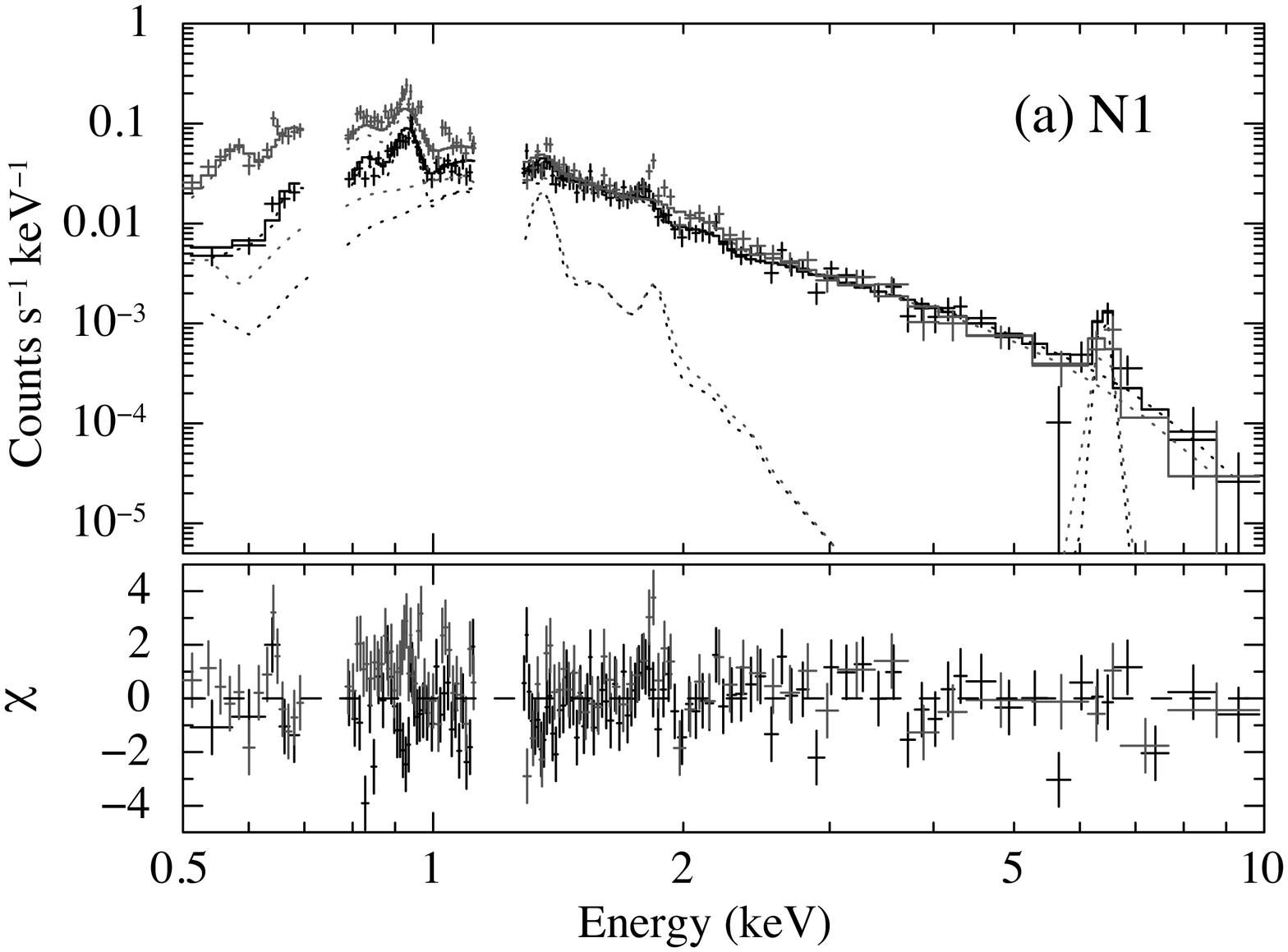}
    \FigureFile(43mm,43mm){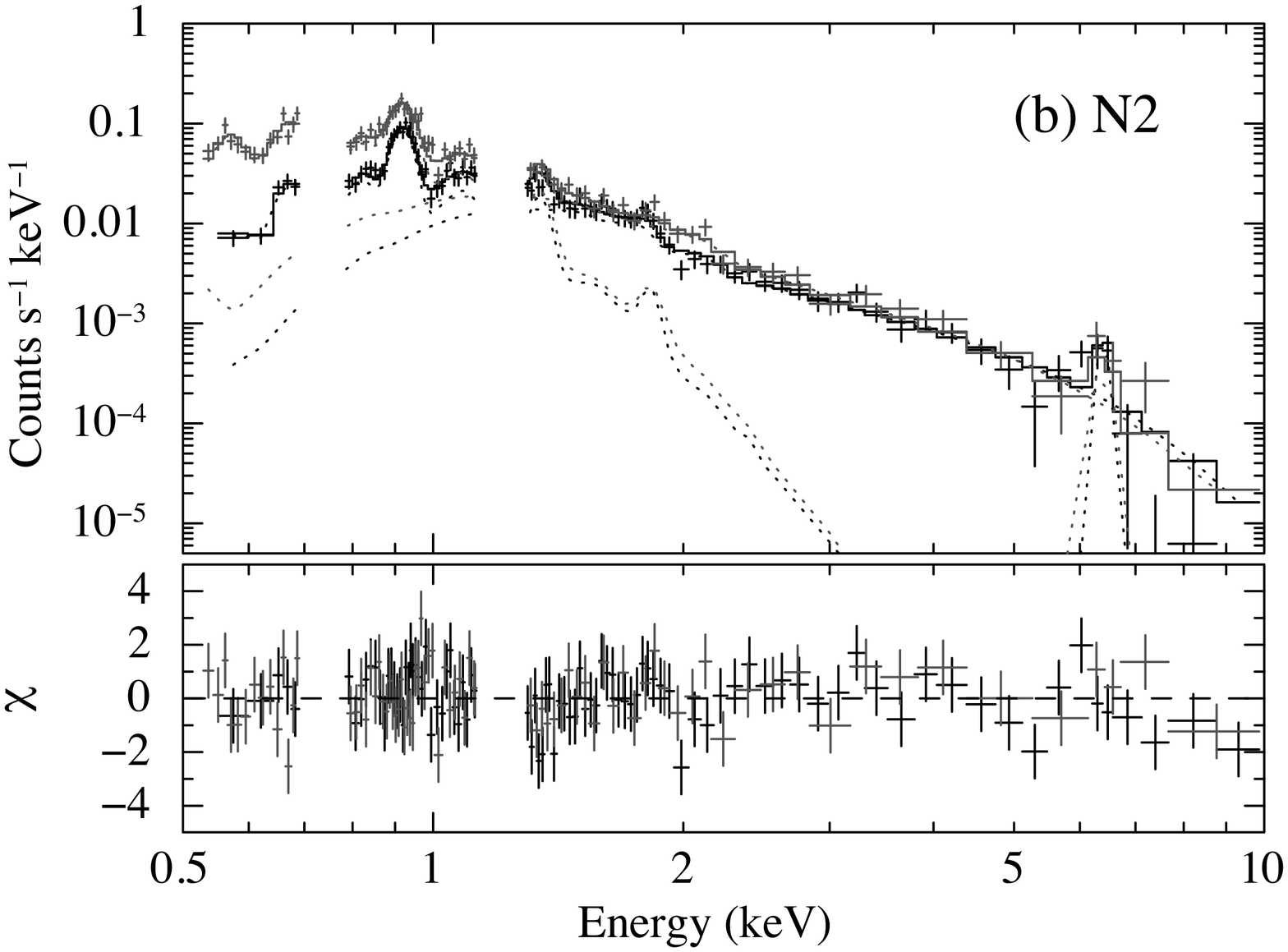}
    \FigureFile(43mm,43mm){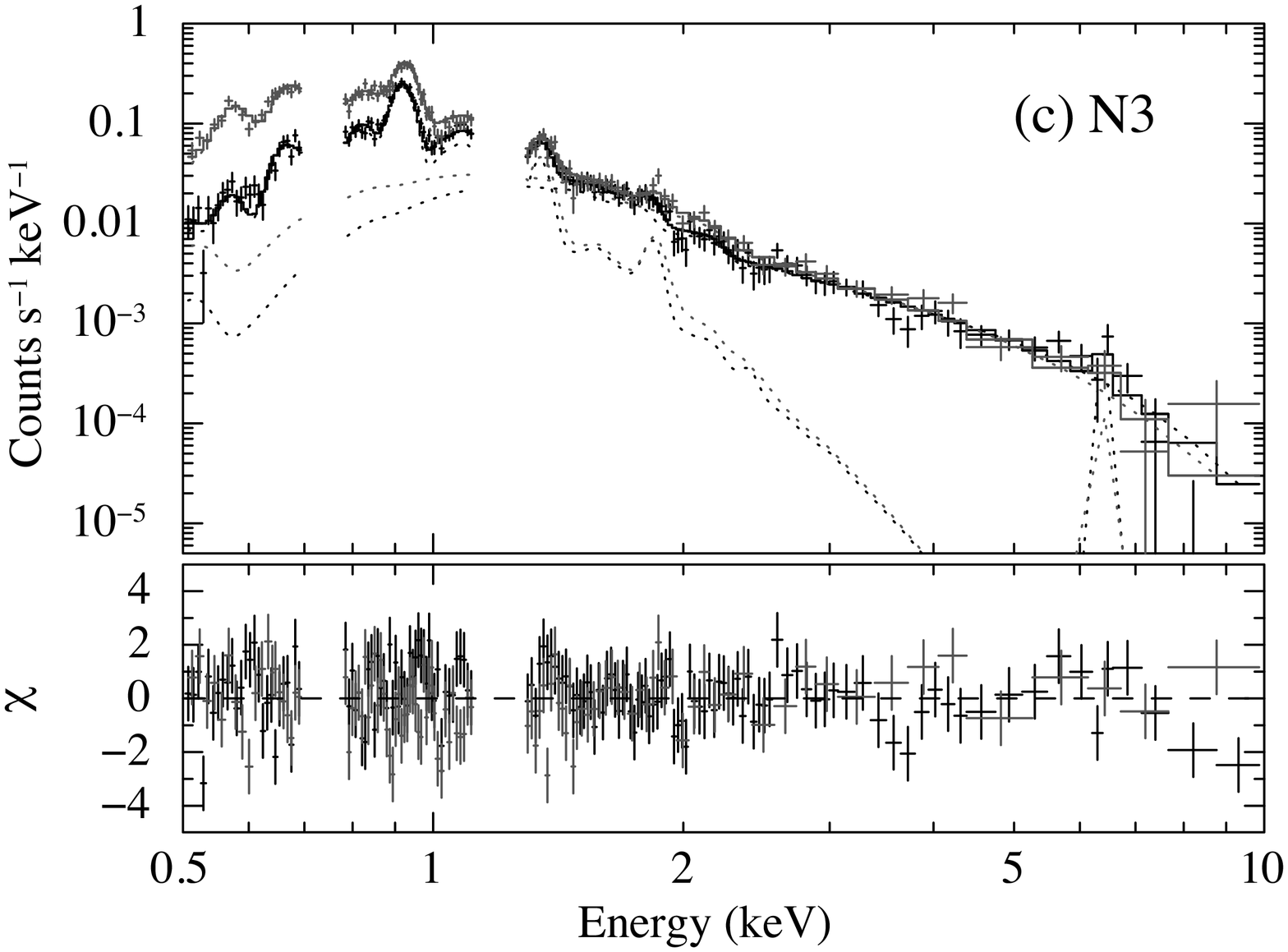}
    \FigureFile(43mm,43mm){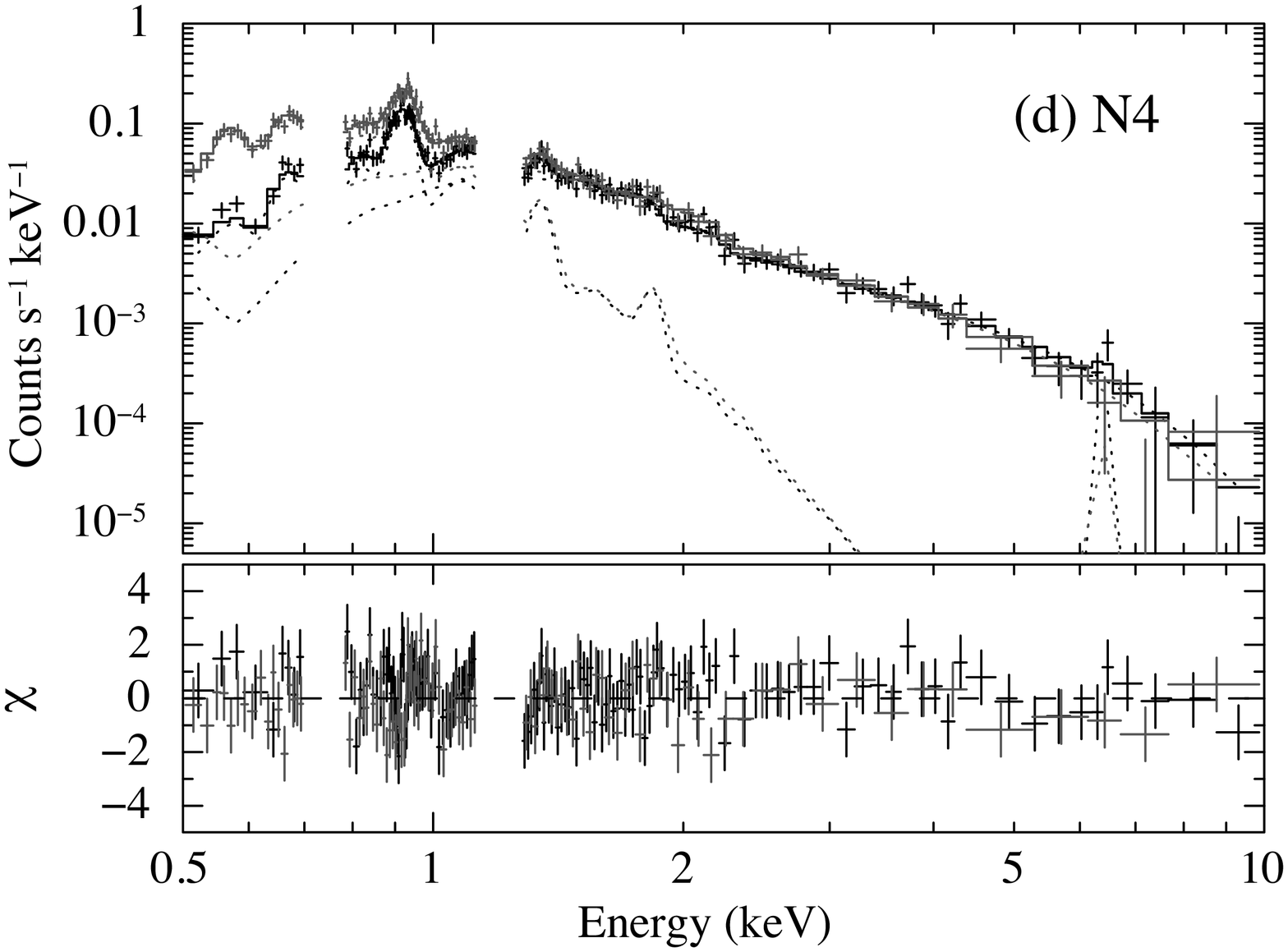}
  \end{center}
  \caption{Same as figure~\ref{fig:spec_w1}, but for the N1 (a), N2 (b), 
  	N3 (c), and N4 (d) regions.}
  \label{fig:spec_n}
\end{figure*}

Figure~\ref{fig:img_n} shows the image of the North FoV. The three colors correspond 
to the same energies as those in figures~\ref{fig:img_all} and \ref{fig:img_w}. 
This region is observed to have a mixture of Balmer-dominated and radiative 
filaments in the optical band (Smith 1997), suggesting that the blast wave is expanding 
into the inhomogeneous ISM. Along the filament from west to east, predominant X-ray 
emission process seems to change from thermal to non-thermal radiation. 
The most remarkable Fe-K emission is observed just behind the soft-thermal filament, 
analogous to the nature of the NE rim region (Yamaguchi et al.\ 2008b). 
In addition, we can see a knotty Fe-K emission located beyond the blast wave boundary. 
The radial profiles of the Soft and Fe-K emissions are shown in figure~\ref{fig:proj_n}.

The spectra were extracted from four regions shown in figure~\ref{fig:img_n}. 
We subtracted the same background as was used in subsection~\ref{ssec:west}. 
The results are shown in figure~\ref{fig:spec_n}. The Fe-K$\alpha$ emission is clearly 
detected in the N1 and N2 regions, and also seen in the N3 and N4 spectra although 
less significant. All the spectra were fitted with the three-component model consisting 
of the soft thermal (\texttt{vpshock}), Fe-K (\texttt{vnei}), and non-thermal emissions. 
Again, the problematic energy ranges (0.70--0.78~keV and 1.13--1.28~keV) were 
ignored in the fitting, and the electron temperature and ionization timescale for 
the \texttt{vnei} component were assumed to be $kT_e = 5.0$~keV and 
$n_e t = 1 \times 10^9$~cm$^{-3}$~s, respectively. 
In the N1 and N2 spectra, all the abundances in the \texttt{vpshock} 
component were fixed to be the solar values. 
The best-fit parameters are given in tables~\ref{tab:n12} and \ref{tab:n34}. 
There is no significant spatial variation in any spectral parameters other than 
the normalization of each component. 
Bocchino et al.\ (2000) argued that the spectra of the North region obtained with 
BeppoSAX and ROSAT can be described by a single temperature NEI model with 
$kT_e \sim 3$~keV. We found, however, that this model failed to reproduce our data, 
but the multi-component model is certainly needed.

\section{Discussion}
\label{sec:discus}

\subsection{Ejecta mass estimation}
\label{ssec:mass}

Since the west rim (W3 region) of RCW~86 is Balmer-dominated (Smith 1997), 
it is likely that the blast wave has been propagated in the low-density and 
relatively uniform ISM. Assuming the emission depth $dl$ of 4.7~pc, 
about one-third of the blast wave radius $R_{\rm bw}$ of 14~pc ($= 17'$), 
we estimate the proton density $n_{\rm H}$ to be 0.3~cm$^{-3}$. Thus, the pre-shock 
ambient density is $n_0 = n_{\rm H}/4 = 0.075$~cm$^{-3}$. Note that when 
the effect of the particle acceleration is unignorable, the value of $n_0$ can be 
smaller, but it does not largely affect the following results.

\begin{figure}[t]
  \begin{center}
    \FigureFile(75mm,75mm){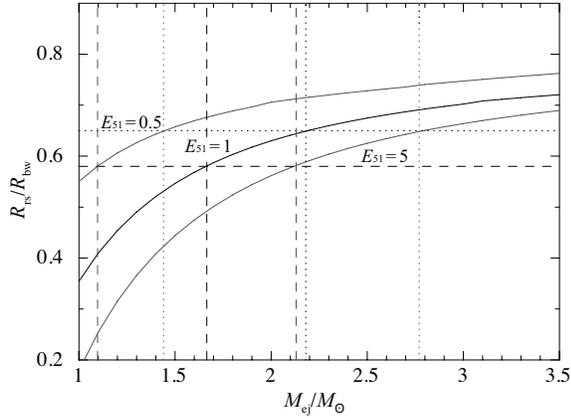}
  \end{center}
  \caption{Ratio of the reverse shock radius ($R_{\rm rs}$) to the blast wave radius 
  ($R_{\rm bw}$) as a function of the ejecta mass ($M_{\rm ej}$), calculated using 
  the formulas of Truelove \& McKee (1999). 
  $E_{51}$ is the explosion energy in the unit of $10^{51}$~erg. The observed values 
  of $R_{\rm rs}/R_{\rm bw}$ in the West and North regions of RCW~86 are indicated as 
  the horizontal dashed and dotted lines, respectively.
  }
  \label{fig:mass}
\end{figure}

Comparing with a numerical calculation of SNRs' dynamical evolution, 
we can estimate the total mass of the ejecta, $M_{\rm ej}$.  
We here simply assume that the SNR ejecta with a constant density profile 
is expanding in an uniform ISM. 
Since the SNR's age is known to be $\sim$1820~yr, the dimensionless age $t^{\ast}$, 
defined by Truelove \& McKee (1999), is obtained as 
\begin{equation}
	t^{\ast} \simeq 1.8~\left(\frac{n_0}{0.075~{\rm cm}^{-3}}\right)^{1/3}
				\left(\frac{E_0}{10^{51}~{\rm erg}}\right)^{1/2} 
				\left(\frac{M_{\rm ej}}{M_{\odot}}\right)^{-5/6},
\end{equation}
where $E_0$ is the explosion energy. 
According to the formulas in table~5 of Truelove \& McKee (1999), the ratio of 
the reverse shock to blast wave radii $R_{\rm rs}/R_{\rm bw}$ can be calculated 
as the function of the ejecta mass. The results for representative explosion energies 
are given in figure~\ref{fig:mass}. 
Assuming that the reverse shock is located at the inner edge of the Fe-K emission 
(W1 region), the reverse shock radius $R_{\rm rs}$ is approximated to be 8~pc 
($= 10'$; see also~figure~\ref{fig:proj_w}), resulting $R_{\rm rs}/R_{\rm bw}$ = 0.58. 
We obtain the the ejecta mass of $\sim$1.1\Msun, $\sim$1.7\Msun, and $\sim$2.1\Msun\  
for the cases of $E_0$ = $5 \times 10^{50}$~erg, $1 \times 10^{51}$~erg, and 
$5 \times 10^{51}$~erg, respectively. 
Differently from the west region, the north rim of the SNR is interacting with a nonuniform 
dense wind-blown shell (Smith 1997; Vink et al.\ 1997). Nevertheless, we try 
a similar investigation to evaluate the uncertainty on the ejecta mass estimation. 
From figure~\ref{fig:proj_n}, the ratio $R_{\rm rs}/R_{\rm bw}$ is obtained to be 0.65. 
This gives the ejecta mass of (1.4--2.8)\Msun, not so far from the result on 
the west region. Investigation into the east rim (Yamaguchi et al.\ 2008b) 
also results the same estimation as the north.

Although the above estimation is derived with a number of assumptions and simplifications 
(i.e., uniform ambient and ejecta structures inside the cavity wall, symmetrical explosion, 
and so on)\footnote{It should also be noted that the recent work by Williams et al.\ (2011) 
suggests that the explosion center of the remnant is offset from the geometric center, 
whereas we assume both are consistent with each other.}$\!$, the estimated ejecta mass 
likely suggests a Type~Ia origin of RCW~86. If this is not the case, at least, the remnant 
originates from a core-collapse explosion of a massive star whose envelope had been 
largely ejected by the pre-explosion stellar wind activity. This is not surprising because 
the SNR is suggested to be expanding in the wind-blown cavity (e.g., Vink et al.\ 1997).

From the integrated spectrum of the entire remnant, we measure the total flux 
of the Fe-K$\alpha$ emission to be $1.7 \times 10^{-4}$~photon~cm$^{-2}$~s$^{-1}$. 
This corresponds to the emission measure $n_e n_{\rm Fe} V$ of 
$4.2 \times 10^{53}$~cm$^{-3}$ ($V$ is the emitting volume) 
for the $kT_e$ = 5~keV case.
Although the density distribution of the Fe-rich plasma is highly uncertain, 
we simply assume that the plasma is uniformly distributed in the spherical shell 
with inner and outer radii of 8~pc and 11~pc, respectively. 
Then, the Fe density and mass are roughly estimated to be 
$n_{\rm Fe}$ = $2 \times 10^{-3}~a^{-0.5}$~cm$^{-3}$ and 
$M_{\rm Fe}$ = $9a^{-0.5}$\Msun, 
respectively, where $a$ is the number of electron per Fe ion. 
Since the value of $a$ is in the range of 1--19 (if the pure-Fe plasma is assumed), 
the Fe mass should be of the order of 1\Msun. 
This estimation would be additional evidence supporting that the remnant 
originates from a Type~Ia supernova rather than a core-collapse supernova 
where the amount of the Fe production is usually less than about 
0.1\Msun\ (e.g., Thielemann et al.\ 1996).

A Type~Ia explosion inside a low-density cavity is quite unusual 
from the observational point of view (Badenes et al.\ 2007). 
However, if the mass accretion rate from the companion star is high enough, 
the accreting white dwarf can blow off a strong optically-thick wind 
(Hachisu et al.\ 1996; 1999). Badenes et al.\ (2007) argued that this outflow 
excavates a large cavity around the progenitor. This can be the case for RCW~86, 
if the SNR really originates from a Type~Ia supernova. 
Alternatively, it is possible that the stellar winds from the other nearby massive stars had 
created the surrounding cavity prior to the explosion, since the remnant is located near 
the OB association (Westerlund 1969; Rosado et al.\ 1996).

\begin{table*}[t]
  \caption{Flux of emission lines.}
  \label{tab:line}         
  \begin{center}
    \begin{tabular}{cccccc}
      \hline
      Region & \multicolumn{3}{c}{Flux ($10^{-6}$ photon cm$^{-2}$ s$^{-1}$)} & 
      	\multicolumn{2}{c}{Flux ratio} \\
	~ & Ar & Ca & Fe & Ar/Fe & Ca/Fe \\
      \hline
	W1 & 0.71 ($<$1.8) & 0.12 ($<$0.90) & $7.0 \pm 0.97$ & $<$0.26 & $<$0.13 \\
	N1 & 0.047 ($<$0.55) & 0.005 ($<$0.38) & $1.4 \pm 0.40$ & $<$0.38 & $<$0.27 \\
	W1+N1 & 0.75 ($<$2.1) & 0.24 ($<$1.1) & $9.0 \pm 1.3$ & $<$0.23 & $<$0.12 \\
  	\hline
    \end{tabular}
  \end{center}
\end{table*}

\subsection{Chemical composition of the ejecta component}
\label{ssec:temperature}

\begin{figure}[t]
  \begin{center}
    \FigureFile(75mm,75mm){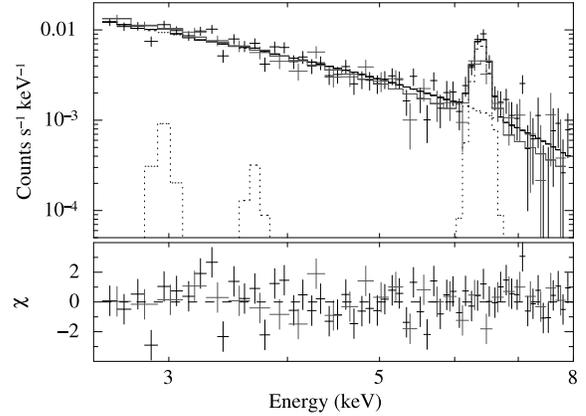}
  \end{center}
  \caption{NXB-subtracted 2.5--8.0~keV spectra where the data from the W1 and N1 
  	regions are integrated. The black and gray represent the FI and BI, respectively. 
	Detection of the Ar (2.96~keV) and Ca (3.69~keV) lines is not significant. 
	}
  \label{fig:spec_merge}
\end{figure}

Whether the origin of RCW~86 is Type~Ia or core-collapse, 
substantial amounts of intermediate-mass elements (i.e., Si and S) 
should be observed similarly to other young SNRs. However, we find 
no evidence for such yields in the ejecta ({\tt vnei}) component. 
Since these elements tend to be laid closer to the contact discontinuity, 
they can be considerably mixed with the shocked ISM, or the electron 
temperature of this layer is too low to emit detectable emission lines. 
On the other hand, higher-$Z$ elements, such as Ar and Ca, are generally synthesized 
in a rather deep layer of Si burning where a part of $^{56}$Ni is also synthesized. 
Therefore, emission from these elements could be detectable. 
Here, we search for K$\alpha$ lines of Ar and Ca in the W1 and N1 spectra,  
where the contribution of non-thermal component is relatively small compared 
to the other rim regions. 
To measure the intensities as precise as possible, we subtract only the NXB 
instead of using the BGD spectrum from the SE FoV, and fit phenomenologically 
the 2.5--8.0~keV spectra with three Gaussians and a power-law. 
The ionization states of both the elements are assumed to be almost neutral, so that
we fix the center energies to be 2.96~keV and 3.69~keV for Ar and Ca, respectively. 
An absorption column of \NH\ = $3.0 \times 10^{21}$~cm$^{-2}$ is assumed, 
but it does not largely affect the X-ray flux in the examined energy band. The line 
fluxes and ratios to that of the Fe-K line we obtained are given in table~\ref{tab:line}. 
Neither Ar nor Ca line is detected significantly, although we have analyzed also 
the integrated spectra of the W1 and N1 regions (figure~\ref{fig:spec_merge}).

\begin{figure}[t]
  \begin{center}
    \FigureFile(70mm,70mm){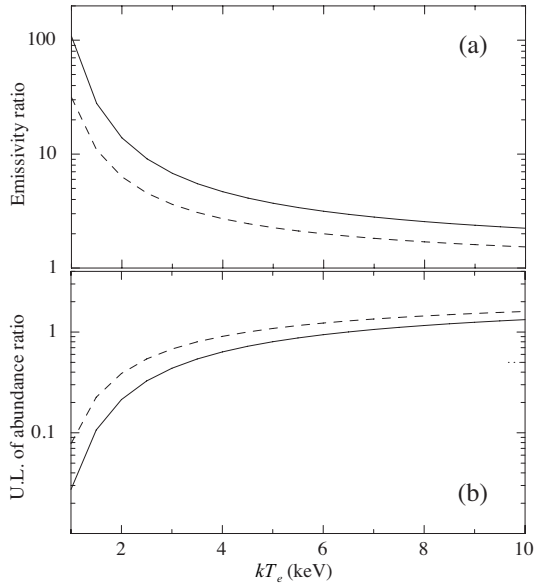}
  \end{center}
  \caption{(a) Emissivity ratios of $\varepsilon_{\rm Ar}$/$\varepsilon_{\rm Fe}$ (solid) 
  	and $\varepsilon_{\rm Ca}$/$\varepsilon_{\rm Fe}$ (dashed) as a function of 
  	the electron temperature. \ 
	(b) Upper limits for the relative abundances of Ar/Fe (solid) and Ca/Fe (dashed) 
	determined from the W1+N1 spectrum.}
  \label{fig:emissivity}
\end{figure}

Using the obtained flux ratios, we can estimate the upper limits for the abundances 
of Ar and Ca relative to Fe. The K-shell line emissivities $\varepsilon_Z (kT_e)$ of 
the element $Z$ is calculated as 
\begin{equation}
\varepsilon_Z (kT_e) \propto \int_{I_Z}^\infty f_{kT_e}(E') \cdot 
\sqrt{E'} \cdot \sigma_Z(E') \cdot \omega_Z \cdot dE',
\end{equation}
where $f_{kT_e}$, $I_Z$, $\sigma_Z$, and $\omega_Z$ are the Maxwell-Boltzmann 
distribution function $2\sqrt{E/[\pi(kT_e)^3]}~$exp$(-E/kT_e)$, the K-shell ionization potential 
and collisional cross section, and the probability of the fluorescent emission, respectively.  
We follow the semi-empirical model by Quarles (1976) for the term of $\sigma_Z$, 
with an assumption that the cross section for the collisional innershell ionization does not 
severely depend on the ionization state in the range from neutral to Ne-like.  
For the values of $\omega_Z$, we use the calculations by Kaastra \& Mewe (1993). 
As the result, the emissivity ratios and upper limits of the relative abundances are 
derived (as a function of the electron temperature) 
as shown in figures~\ref{fig:emissivity}a and \ref{fig:emissivity}b, respectively. 
Note that the latter is determined from the merged W1+N1 spectrum. 
Unfortunately, the electron temperature of the ejecta component cannot be constrained 
by the spectral analysis, because the hard X-ray continuum of this SNR seems to be 
dominated by the non-thermal radiation. However, given that most young SNRs have 
an electron temperature of less than 10~keV, we can reject the possibility 
that the relative abundances of Ar/Fe and Ca/Fe are larger than 1.6~solar. 
This is not incompatible with the predicted yields for Type~Ia supernovae.

\subsection{Possible evidence of the ejecta knot}
\label{ssec:knot}

In the north region, we have observed the Fe-K emission visible beyond 
the blast wave filament (the N2 region in Figure~\ref{fig:img_n}). 
This feature is quite distinct from those of the other ejecta distribution in this SNR; 
the Fe-rich ejecta are usually observed inside the main shell that is dominated 
by the soft thermal emission. Therefore, different interpretation for its origin should be 
considered. Here we analogize it with ejecta bullets associated with the Tycho SNR 
or the Vela SNR, the metal-rich clumps located outside the blast-shock boundary 
(e.g., Tycho: Vancura et al.\ 1995; Decourchelle et al.\ 2001; 
Vela: Aschenbach et al.\ 1995; Miyata et al.\ 2001). 
Since the optical observations of RCW~86 had suggested that the blast wave in 
the North region is expanding into the dense inhomogeneous ISM (Smith 1997), 
it is likely that the shock had decelerated very recently, resulting that the ejecta knot 
overran the primary blast wave. If this is the case, a bow-shock feature can be observed 
at the head of the knot, similarly to the Vela shrapnels. However, the statistics of our data 
is too poor to investigate such detailed structure. Deeper observation will help reveal 
the origin of this peculiar Fe-K emission.

\section{Summary}
\label{sec:summary}

We have presented the first detection of Fe-K emission from the west, north, 
and south regions of the SNR RCW~86. Detailed imaging and spectral analysis 
have been performed for the former two regions. 
The spatial distributions of the Fe-K$\alpha$ and hard X-ray emissions are not 
correlated with each other, rejecting the possibility that the Fe-K$\alpha$ emission 
originates from the fluorescence by supra-thermal electrons or hard X-rays. 
In the west rim, the Fe-K emission is apparently enhanced at the inward region 
with respect to the forward shock, which supports its origin of supernova ejecta 
heated by the reverse shock. From the derived ambient density (0.075~cm$^{-3}$) 
and radii of the forward and reverse shocks (14~pc and 8~pc, respectively), 
the total ejecta mass is roughly estimated to be (1--2)\Msun. 
This suggests that the SNR originates from a Type~Ia supernova explosion 
rather than a core-collapse one considered previously (e.g., Rosado et al.\ 1996). 
The total mass of Fe, of the order of 1\Msun, also supports its Type~Ia origin. 
The low-density cavity surrounding the Type~Ia progenitor is not usual. 
It could be formed by so-called accretion winds from the progenitor itself or 
the outflow from the other nearby OB stars. 
In the north rim, we find another Fe-K emission located beyond the primary blast wave 
boundary, which is analogous to the ejecta fragments observed in the Tycho and Vela SNRs. 
To confirm its nature more conclusively, a future deep observation 
is required. No significant evidence for low-ionized Ar and Ca emissions is observed. 
This constrains the relative abundances of Ar/Fe and Ca/Fe to be less than $\sim 1.6$~solar.

\bigskip

{\it Note}---After the first submission of this paper, the results on the same object 
had been reported by Williams et al.\ (2011; arXiv:1108.1207). 
They presented the infrared and X-ray observations and arrived at 
a similar interpretation, a Type~Ia origin of this remnant.

\bigskip

The authors deeply appreciate a number of constructive suggestions from the 
anonymous referee. We also thank to Drs.\ Adam Foster and Randall K.\ Smith 
for providing useful information about AtomDB, and Drs.\ Patrick O.\ Slane and 
Daniel Patnaude for stimulating discussion. Dr.\ Midori Ozawa largely contributed 
to planning the mapping observation of RCW~86 using Suzaku.
H.Y.\ and H.U.\ are supported by Japan Society for the Promotion of Science (JSPS) 
Research Fellowship for Research Abroad and Young Scientists, respectively. 
K.K.\ is supported by the Challenging Exploratory Research program
(No.~20654019) of the Ministry of Education, Culture, Sports, Science
and Technology (MEXT).


\begin{thebibliography}{}

\bibitem[Aharonian et al.(2009)]{2009ApJ...692.1500A} Aharonian, F., et 
al.\ 2009, \apj, 692, 1500 

\bibitem[Anders \& Grevesse(1989)]{1989GeCoA..53..197A} Anders, E., \& 
Grevesse, N.\ 1989, \gca, 53, 197 

\bibitem[Aschenbach et al.(1995)]{1995Natur.373..587A} Aschenbach, B., 
Egger, R., \& Tr{\"u}mper, J.\ 1995, \nat, 373, 587 

\bibitem[Badenes et al.(2007)]{2007ApJ...662..472B} Badenes, C., Hughes, 
J.~P., Bravo, E., \& Langer, N.\ 2007, \apj, 662, 472 

\bibitem[Bamba et al.(2000)]{2000PASJ...52.1157B} Bamba, A., Koyama, K., \& 
Tomida, H.\ 2000, \pasj, 52, 1157 

\bibitem[Bocchino et al.(2000)]{2000A&A...360..671B} Bocchino, F., Vink, 
J., Favata, F., Maggio, A., \& Sciortino, S.\ 2000, \aap, 360, 671 

\bibitem[Borkowski et al.(2001a)]{2001ApJ...548..820B} Borkowski, K.~J., 
Lyerly, W.~J., \& Reynolds, S.~P.\ 2001a, \apj, 548, 820 

\bibitem[Borkowski et al.(2001b)]{2001ApJ...550..334B} Borkowski, K.~J., 
Rho, J., Reynolds, S.~P., \& Dyer, K.~K.\ 2001b, \apj, 550, 334 

\bibitem[Dickel et al.(2001)]{2001ApJ...546..447D} Dickel, J.~R., Strom, 
R.~G., \& Milne, D.~K.\ 2001, \apj, 546, 447 

\bibitem[Decourchelle et al.(2001)]{2001A&A...365L.218D} Decourchelle, A., 
Sauvageot, J.~L., Audard, M., et al.\ 2001, \aap, 365, L218 

\bibitem[Green (2009)]{green} Green, D.~A., 2009, `A Catalogue of Galactic Supernova 
Remnants (2009 March version)', Astrophysics Group, Cavendish Laboratory, 
Cambridge, United Kingdom $<$http://www.mrao.cam.ac.uk/surveys/snrs/$>$

\bibitem[Hachisu et al.(1996)]{1996ApJ...470L..97H} Hachisu, I., Kato, M., 
\& Nomoto, K.\ 1996, \apjl, 470, L97 

\bibitem[Hachisu et al.(1999)]{1999ApJ...522..487H} Hachisu, I., Kato, M., 
\& Nomoto, K.\ 1999, \apj, 522, 487 

\bibitem[Helder et al.(2009)]{2009Sci...325..719H} Helder, E.~A., et al.\ 
2009, Science, 325, 719 

\bibitem[van der Heyden et al.(2002)]{2002A&A...392..955V} van der Heyden, K.~J., 
Behar, E., Vink, J., Rasmussen, A.~P., Kaastra, J.~S., Bleeker, J.~A.~M., Kahn, S.~M., 
\& Mewe, R.\ 2002, \aap, 392, 955 

\bibitem[Hwang et al.(2002)]{2002ApJ...581.1101H} Hwang, U., Decourchelle, 
A., Holt, S.~S., \& Petre, R.\ 2002, \apj, 581, 1101 

\bibitem[Iwamoto et al.(1999)]{1999ApJS..125..439I} Iwamoto, K., Brachwitz, 
F., Nomoto, K., Kishimoto, N., Umeda, H., Hix, W.~R., 
\& Thielemann, F.-K.\ 1999, \apjs, 125, 439 

\bibitem[Kaastra \& Mewe(1993)]{1993A&AS...97..443K} Kaastra, J.~S., \& Mewe, R.\ 
1993, \aaps, 97, 443 

\bibitem[Kallman et al.(2004)]{2004ApJS..155..675K} Kallman, T.~R., 
Palmeri, P., Bautista, M.~A., Mendoza, C., 
\& Krolik, J.~H.\ 2004, \apjs, 155, 675 

\bibitem[Katsuda et al.(2011)]{2011ApJ...730...24K} Katsuda, S., et al.\ 
2011, \apj, 730, 24 

\bibitem[Koyama et al.(1995)]{1995Natur.378..255K} Koyama, K., Petre, R., 
Gotthelf, E.~V., Hwang, U., Matsuura, M., Ozaki, M., \& Holt, S.~S.\ 1995, 
\nat, 378, 255 

\bibitem[Koyama et al.(2007a)]{2007PASJ...59S.245K} Koyama, K., et al.\ 
2007a, \pasj, 59, S245 

\bibitem[Koyama et al.(2007b)]{2007PASJ...59S..23K} Koyama, K., et al.\ 
2007b, \pasj, 59, S23 

\bibitem[Lallement(2004)]{2004A&A...422..391L} Lallement, R.\ 2004, \aap, 422, 391 

\bibitem[Makishima(1986)]{1986LNP...266..249M} Makishima, K.\ 1986, The 
Physics of Accretion onto Compact Objects, ed.\ Mason, K.~P., Watson, M.~G., \& 
White, N.~E.\ (Berlin: Springer), 266, 249 

\bibitem[Miyata et al.(2001)]{2001ApJ...559L..45M} Miyata, E., Tsunemi, H., 
Aschenbach, B., \& Mori, K.\ 2001, \apjl, 559, L45 

\bibitem[Nomoto et al.(1984)]{1984ApJ...286..644N} Nomoto, K., Thielemann, 
F.-K., \& Yokoi, K.\ 1984, \apj, 286, 644 

\bibitem[Pisarski et al.(1984)]{1984ApJ...277..710P} Pisarski, R.~L., 
Helfand, D.~J., \& Kahn, S.~M.\ 1984, \apj, 277, 710 

\bibitem[Quarles(1976)]{1976PhRvA..13.1278Q} Quarles, C.~A.\ 1976, \pra, 
13, 1278 

\bibitem[Rho et al.(2002)]{2002ApJ...581.1116R} Rho, J., Dyer, K.~K., 
Borkowski, K.~J., \& Reynolds, S.~P.\ 2002, \apj, 581, 1116 

\bibitem[Rosado et al.(1996)]{1996A&A...315..243R} Rosado, M., 
Ambrocio-Cruz, P., Le Coarer, E., \& Marcelin, M.\ 1996, \aap, 315, 243 

\bibitem[Smith(1997)]{1997AJ....114.2664S} Smith, R.~C.\ 1997, \aj, 114, 
2664 

\bibitem[Tamagawa et al.(2009)]{2009PASJ...61S.167T} Tamagawa, T., et al.\ 
2009, \pasj, 61, S167 

\bibitem[Tamura et al.(2009)]{2009ApJ...705L..62T} Tamura, T., et al.\ 
2009, \apjl, 705, L62 

\bibitem[Thielemann et al.(1996)]{1996ApJ...460..408T} Thielemann, F.-K., 
Nomoto, K., \& Hashimoto, M.-A.\ 1996, \apj, 460, 408 

\bibitem[Tomida et al.(1999)]{1999AN....320..342T} Tomida, H., Koyama, K., 
\& Yamauchi, S.\ 1999, Astron. Nachr. 320, 342 

\bibitem[Truelove \& McKee(1999)]{1999ApJS..120..299T}
Truelove, J.~K., \& McKee, C.~F.\ 1999, \apjs, 120, 299

\bibitem[Ueno et al.(2007)]{2007PASJ...59S.171U} Ueno, M., et al.\ 2007, 
\pasj, 59, S171 

\bibitem[Uchiyama et al.(2009)]{2009PASJ...61S...9U} Uchiyama, H., et al.\ 
2009, \pasj, 61, S9 

\bibitem[Vancura et al.(1995)]{1995ApJ...441..680V} Vancura, O., 
Gorenstein, P., \& Hughes, J.~P.\ 1995, \apj, 441, 680 

\bibitem[Vink et al.(1997)]{1997A&A...328..628V} Vink, J., Kaastra, J.~S., 
\& Bleeker, J.~A.~M.\ 1997, \aap, 328, 628 

\bibitem[Vink et al.(2006)]{2006ApJ...648L..33V} Vink, J., Bleeker, J., van 
der Heyden, K., Bykov, A., Bamba, A., \& Yamazaki, R.\ 2006, \apj, 648, 
L33 

\bibitem[Warren \& Hughes(2004)]{2004ApJ...608..261W} Warren, J.~S., \& 
Hughes, J.~P.\ 2004, \apj, 608, 261 

\bibitem[Westerlund(1969)]{1969AJ.....74..879W} Westerlund, B.~E.\ 1969, 
\aj, 74, 879 

\bibitem[Whiteoak \& Green(1996)]{1996A&AS..118..329W} Whiteoak, J.~B.~Z., \& 
Green, A.~J.\ 1996, \aaps, 118, 329 

\bibitem[Wilms et al.(2000)]{2000ApJ...542..914W} Wilms, J., Allen, A., 
\& McCray, R.\ 2000, \apj, 542, 914 

\bibitem[Yamaguchi et al.(2008a)]{2008PASJ...60S.141Y} Yamaguchi, H., et 
al.\ 2008a, \pasj, 60, S141 

\bibitem[Yamaguchi et al.(2008b)]{2008PASJ...60S.123Y} Yamaguchi, H., 
Koyama, K., Nakajima, H., Bamba, A., Yamazaki, R., Vink, J., 
\& Kawachi, A.\ 2008b, \pasj, 60, S123 

\end{thebibliography}
\end{document}